\documentclass[aps,prd,11pt,preprint,nofootinbib,showkeys,unsortedaddress]{revtex4-2}
\usepackage{amsmath,amssymb,amsthm,mathrsfs,bbm}
\usepackage{latexsym,amscd,amsbsy,amsfonts,dsfont}
\usepackage{graphicx}
\usepackage[utf8]{inputenc}
\usepackage{subfigure}  
\usepackage{float}
\usepackage{slashed}
\usepackage{cancel}
\usepackage{multirow}
\usepackage{soul}
\usepackage{physics}
\usepackage{comment}
\usepackage[usenames,dvipsnames]{xcolor}
\usepackage[
      colorlinks=true,
      linktocpage=true,
      breaklinks=true,
      linkcolor=Aquamarine,
      urlcolor=Purple,
      filecolor=black,
      citecolor=Purple,
      menucolor=black,
      pdfstartview=FitV,
      bookmarksopen=true
      ]{hyperref}

\definecolor{emerald}{rgb}{0.31, 0.78, 0.47}

\makeindex

\begin{document}

\title{Chronology Protection Implementation in Analogue Gravity}

\author{Carlos Barcel\'o}
\email{carlos@iaa.es}
\affiliation{Instituto de Astrof\'{\i}sica de Andaluc\'{\i}a (IAA-CSIC), Glorieta de la Astronom\'{\i}a, 18008 Granada, Spain}
\author{Jokin Eguia S\'anchez}
\affiliation{Department of Cell Biology and Histology, Faculty of Medicine and Nursing, University of the Basque Country (UPV/EHU), Barrio Sarriena S/N, 48940 Leioa, Spain}
\author{Gerardo Garc\'ia-Moreno}
\email{ggarcia@iaa.es}
\affiliation{Instituto de Astrof\'{\i}sica de Andaluc\'{\i}a (IAA-CSIC), Glorieta de la Astronom\'{\i}a, 18008 Granada, Spain}
\author{Gil Jannes}
\email{gjannes@ucm.es}
\affiliation{Department of Financial and Actuarial Economics \& Statistics, Universidad Complutense de Madrid, Campus Somosaguas s/n, 28223 Pozuelo de Alarc\'on (Madrid), Spain}

\begin{abstract}
{Analogue gravity systems offer many insights into gravitational phenomena, both at the classical and at the semiclassical level. The existence of an underlying Minkowskian structure (or Galilean in the non-relativistic limit) in the laboratory has been argued to directly forbid the simulation of geometries with Closed Timelike Curves (CTCs) within analogue systems. We will show that this is not strictly the case. In principle, it is possible to simulate spacetimes with CTCs whenever this does not entail the presence of a chronological horizon separating regions with CTCs from regions that do not have CTCs. We find an Analogue-gravity Chronology protection mechanism very similar in spirit to Hawking's Chronology Protection hypothesis. We identify the universal behaviour of analogue systems near the formation of such horizons and discuss the further implications that this analysis has from an emergent gravity perspective. Furthermore, we build explicit geometries containing CTCs, for instance spacetimes constructed from two warp-drive configurations, that might be useful for future analysis, both from a theoretical and an experimental point of view. }
\end{abstract}


\maketitle

{

\hypersetup{citecolor=black, linkcolor=black,urlcolor=black}

\tableofcontents

}

\section{Introduction}
\label{Sec:Introduction}

It is by now well known that many systems akin to condensed matter systems, in the sense of being composed by a large amount of elementary building blocks (atoms or abstract particles), exhibit a behaviour in certain regimes which can be characterized by the presence of some effective fields, classical or quantum, moving in an effectively curved Lorentzian geometry. These behaviours are collectively called ``analogue gravity"~\cite{Barcelo2011,Volovik2009}. In its broadest description, the analogue gravity program intends to obtain new insights into gravitational behaviours by analyzing their equivalent counterparts within these analogue frameworks. The reverse direction: acquiring new ideas about laboratory systems by importing gravitational notions and techniques, is also part of the analogue gravity realm. 

The most paradigmatic analysis within this program has been the theoretical and experimental verification of Hawking radiation within black hole configurations even when these take place in the context of an effective and collective phenomenon. The appearance of spontaneous Hawking radiation in Bose-Einstein condensates has been observed in~\cite{Steinhauer2021} as originally suggested in~\cite{Garay1999}. Apart from black holes, since the late 90s, many other types of geometries have also been proposed in different laboratory settings~\cite{Barcelo2011}, such as rotating geometries~\cite{Jannes2015,Weinfurtner2016,Faccio2021}, cosmological solutions~\cite{Barcelo2003} with a very recent experimental realization with fluids of light~\cite{Steinhauer2021b}, anti-de Sitter spacetime~\cite{Barcelo2002}, and even warp-drive geometries~\cite{Fischer2002,Finazzi2011}. For a more exhaustive list, see~\cite{Barcelo2011} and references therein. 

In this paper we are interested precisely in some puzzles that appear when playing with these warp drive geometries. It is well known that one can in principle use warp drives to build time machines~\cite{Everett1995}. This raises the question whether one could simulate a geometry with Closed Timelike Curves (CTCs) within an analogue system inspired by warp drives. The laboratory systems used to build analogue gravity configurations are always embedded in our locally Minkowskian world. However, the Minkowski structure of our current fundamental description of Nature typically does not even play a role in the analogue geometries: these geometries can be obtained directly within a Galilean description of the laboratory. In both cases, it seems that the causal structure of the background (Minkowski or Galilean) prohibits the simulation of causally pathological spacetimes in the embedded analogue gravity system~\cite{Visser1997b}. However, by analyzing different situations, we will show that this is in fact not strictly the case. The simulation of spacetimes with CTCs \emph{per se} does not present insurmountable obstacles. The real problem appears when the relevant spacetimes posses a chronological horizon~\cite{Hawking1973}, that is, a surface separating a region with CTCs from another with a standard causality. For external or laboratory observers, the inability to generate a region with CTCs manifests itself in the form of divergences in certain properties of the local physics that they experience. On the other hand, for an internal observer without direct access to the underlying causal structure, the inability to produce CTCs manifests itself through some effective protection mechanism. As we will discuss, these mechanisms are reminiscent of Hawking's Chronology Protection Conjecture. 

Throughout our discussion, we will revise several simple configurations of spacetimes with CTCs, such as warp drive spacetimes, G\"odel spacetime and Misner spacetime. We present versions of these geometries that are amenable to further analysis both in the context of analogue gravity but also from a purely geometrical perspective. In order to maintain an explicit connection with potential laboratory realizations of such CTC spacetimes, we will focus on a concrete substratum, namely a generalized Bose-Einstein condensate with anisotropic masses~\cite{Barcelo2001,Barcelo2011}. The inverse acoustic metric for linear perturbations on this quantum fluid can be written
\begin{equation}
    g^{\mu\nu}=\frac{\mu}{\rho_0 c} \left(
    \begin{array}{c|c}
    -1 & -v^i \\
    \hline \\
    -v^j & c^2 h^{ij}-v^i v^j
    \end{array}
    \right).
        \label{Eq:AcousticMetric}
\end{equation}
Here $\rho_0$ is the background fluid density, $c$ the local speed of sound and $v^{i}(x)$ the velocity of the fluid, which is simply the gradient of the phase of the macroscopic BEC wave function. The matrix $h_{ij}$ (or its inverse $h^{ij}$) takes into account any anisotropy acquired by the effective masses of the bosons in the condensate: $m_{ij}=\mu h_{ij}$, where $\mu$ is an arbitrary conformal constant. In simple (isotropic) BECs, $h_{ij}$ is just a multiple of the identity matrix. For a weakly interacting BEC there is also the following relation between $c, \mu$ and the effective coupling constant of the condensate $\lambda$: $c^2=\lambda \rho/\mu$.

The metric~\eqref{Eq:AcousticMetric} automatically inherits the stable causality property~\cite{Hawking1973,Wald1984} from the background structure~\cite{Visser1997b}. In particular, it contains a globally defined time function $t$ since 
\begin{equation}
    g^{\mu \nu} \partial_{\mu} t \partial_{\nu} t = - \frac{\mu}{\rho_0 c} \leq 0.
\end{equation}

This appears to automatically rule out the possibility of simulating metrics containing CTCs. As stated above, we will see that without further qualification this is in fact not strictly true. Moreover, in the cases in which we really find an obstruction, it is interesting to analyze when and how these effective-metric descriptions break down, and how these breakdowns are related to mechanisms of Chronology protection. 

A brief outline of the remainder of this work is the following. We begin in Section~\ref{Sec:Godel} with the warm-up exercise of trying to simulate a G\"odel spacetime and mild deformations thereof in the system described above. We will find that, although it is possible to simulate CTCs, they are trivial in a sense that we will specify concretely. Furthermore, we will find that it is impossible to simulate a modification of G\"odel's spacetime such that a chronologically well-behaved region with no CTCs evolves into a region with CTCs, due to the divergence of the speeds of the fluid required. Motivated by this exercise, we try to analyze whether this is a generic feature of spacetimes containing CTCs. For that purpose, we introduce in Section~\ref{Sec:Survey} a catalogue of geometries amenable to simulation in analogue gravity. Some of them do not have a General Relativistic counterpart. In Subsection~\ref{Subsec:CTC-warp drives} we describe spacetimes containing CTCs engineered from two warp-drive bubbles. We discuss the impossibility of doing it in $1+1$ spacetime dimensions, with special emphasis on the point that CTCs in such dimensionality require non-trivial topologies. Based on these warp drive tube geometries, we introduce a family of simpler geometries which are qualitatively similar to them but much easier to handle in Subsection~\ref{Subsec:Extended-Warpdrives}. We conclude Section~\ref{Sec:Survey} with a discussion of $1+1$-dimensional spacetimes in Subsection~\ref{Subsec:Misner}. We introduce the archetypal example of a spacetime containing a chronological horizon, Misner's spacetime, and then discuss how an eternal cylinder with a flat metric can be understood as having ``trivial" CTCs by a simple interchange of the time and space coordinates. A reader interested just in the Chronology Protection mechanism in Analogue gravity can safely skip these first sections and jump directly to Section~\ref{Sec:Analogue-simulations}, which contains a detailed description of the possibility of simulating trivial CTCs in our analogue model, and the impossibility of simulating chronological horizons. Furthermore, we identify the insurmountable difficulty that every standard analogue gravity model would face when trying to simulate a chronological horizons. In Section~\ref{Sec:Chronology-Protection} we discuss the interplay of our analysis and Hawking's Chronology Protection conjecture, its implications for the emergent gravity program and we also connect with recent related discussions in the literature. Finally, we finish in Section~\ref{Sec:Summary} by summarizing the content of the article and describing potential directions for future work. 

\textit{Notation and conventions.} We will use the signature $(-,+,...,+)$ for the spacetime metric and follow the Misner-Thorne-Wheeler conventions for the curvature tensors~\cite{Misner1974}. Greek indices $(\mu, \nu, ...)$ will run from $0$ to $D$, representing spacetime indices, whereas latin indices $(i,j...)$ will run from $1$ to $D$ and represent spatial indices. Einstein's summation convention is used throughout the work unless otherwise stated.

\section{Attempts to simulate G\"odel spacetime}
\label{Sec:Godel}

As a warm-up exercise we will describe G\"odel's metric as the archetypal example of a geometry which contains CTCs~\cite{Godel1949,Hawking1973}. The purpose of this section is twofold. First, we will describe the geometric properties of G\"odel's spacetime. Many of these properties will be shared by any spacetime containing CTCs, thus allowing us to focus the discussion on the essential features for any successful simulation of CTCs in an analogue model. Second, we will dig into the problems that appear when one attempts to simulate such chronologically pathological spacetimes. A more general discussion concerning generic spacetimes displaying CTCs will be provided later. The starting point of this section has a substantial overlap with the unpublished work~\cite{Volovik}. An analysis similar to the one presented here for the simulation of G\"odel geometry in an optic system was presented in~\cite{Fiorini2021}. The identification of the metric components with the physical parameters of the analogue system do not seem to be done in the correct way, and hence the divergences that we observe in the horizons here are absent.

G\"odel's spacetime is a solution of the Einstein equations with suitable sources, namely a negative cosmological constant $\Lambda$ and the energy momentum tensor of a pressureless perfect fluid with a density: $\rho \propto - \Lambda$. In appropriate coordinates $(t,r,\phi,z)$, it can be written as follows~\cite{Kajari2004}: 
\begin{equation}
    ds^2 = -dt^2 + \frac{dr^2}{1 + \frac{r^2}{4 \omega^2}} + r^2 \left( 1 - \frac{r^2}{4 \omega^2}\right) d \phi^2 + dz^2 - \frac{\sqrt{2}}{\omega} r^2 dt d \phi 
    \label{Eq:Godel}
\end{equation}
where $\omega$ is a parameter characterizing the solution and related to the density and hence also trivially related to the cosmological constant as $\omega^2 = - \Lambda$. That this geometry contains CTCs can be seen as follows: for $r \geq r_C=2\omega$, the (Killing) vector field $\partial_{\phi}$ becomes timelike. Since such a vector needs to be periodically identified to avoid a conical singularity at $r=0$, we have that $\phi \sim \phi + 2 \pi$. Hence, this vector field has closed orbits. Since it becomes timelike at $r \geq r_C$, it is trivial to conclude that the orbits of $\phi$ for $r>r_C$ are CTCs. 

It seems that no CTCs pass through the region $r < r_C$. However, we must take into account that this geometry is completely homogeneous, in fact it contains a group of five Killing vector fields acting transitively on the manifold~\cite{Hawking1973}. This means that every point of the manifold can be mapped by a symmetry transformation to any other point on the manifold. Hence, CTCs pass through every single point in this spacetime. However, there are no CTCs confined to the region $r<r_C$, in fact every CTC passing through the region $r<r_C$ crosses the cylinder $r = r_C$ an even number of times. 

From the point of view of an acoustic metric, we can realize that this precise system of coordinates allows for a direct realization of G\"odel's metric with suitable fluid parameters.  The non-vanishing components of the inverse G\"odel metric in these coordinates are
\begin{align}
& g^{tt}= -F(r), \nonumber \\
& g^{t\phi}= -\frac{1}{\sqrt{2}\omega }    \frac{ 1 }{\left( 1-\frac{r^2 }{4\omega^2} \right) } F(r), \nonumber \\
& g^{\phi\phi}=\frac{1}{r^2 }    \frac{ 1}{\left( 1- \frac{r^2}{ 4\omega^2} \right) } F(r), \nonumber \\
& g^{rr}= \frac{ \left(1+ \frac{r^2}{4\omega^2} \right)^2 }{ \left(1- \frac{r^2}{4\omega^2} \right)} F(r), \nonumber \\
& g^{zz}= \frac{ \left(1+ \frac{r^2}{4\omega^2} \right) }{\left(1- \frac{r^2}{4\omega^2} \right)}  F(r).
\label{Eq:InverseGodel}
\end{align}
with $F(r)$ defined as
\begin{equation}
 F= \frac{ \left(1- \frac{r^2 } {4\omega^2} \right) }{ \left(1+ \frac{r^2 }{ 4\omega^2} \right)}.
\end{equation}

Comparing Eq.~\eqref{Eq:InverseGodel} with Eq.~\eqref{Eq:AcousticMetric} we realize that we simply need a motion of the fluid in the $\phi$-direction $v^{i} = v_{\phi} \delta^{i}_{\ \phi}$. Taking into account the change of coordinates to a cylindrical-like coordinate system we have that the azimuthal velocity must be
\begin{equation}
    v_{\phi} = \frac{1}{\sqrt{2} \omega} \frac{r}{1 - \frac{r^2}{4 \omega^2}}.
\end{equation}
On the other hand the three principal directions of the anisotropy matrix $h^{ij}$ need to obey
\begin{align}
    c^2 h^{rr}= \frac{ \left( 1 + \frac{r^2}{4 \omega^2}\right)^2}{\left(1 -\frac{r^2}{4 \omega^2} \right)}, \nonumber \\
    c^2 h^{\phi\phi} = \frac{\left(1 + \frac{r^2}{4 \omega^2}\right)}{\left(1 - \frac{r^2}{4 \omega^2}\right)^2}, \nonumber \\
    c^2 h^{zz}= \frac{\left(1 + \frac{r^2}{4 \omega^2}\right)}{\left(1 - \frac{r^2}{4 \omega^2}\right)}.
    \label{Eq:Principal_directions}
\end{align}
These quantities can be interpreted as three anisotropic sound speeds $c_r^2, c_{\phi}^2$ and $c_z^{2}$, respectively. In addition, we notice that for the weakly interacting BEC we have $\lambda/c^3=F$, or equivalently
\begin{equation}
   c^2= \lambda^{2/3} \frac{ \left(1 + \frac{r^2}{4 \omega^2} \right)^{2/3}} {\left(1 - \frac{r^2}{4 \omega^2} \right)^{2/3}},
\end{equation}
we finally obtain  
\begin{align}
    h^{rr}= \lambda^{-2/3}   \frac{ \left( 1 + \frac{r^2}{4 \omega^2}\right)^{4/3}}{\left(1 -\frac{r^2}{4 \omega^2} \right)^{1/3}}, \nonumber \\
    h^{\phi\phi} = \lambda^{-2/3}   \frac{\left(1 + \frac{r^2}{4 \omega^2}\right)^{1/3}}{\left(1 - \frac{r^2}{4 \omega^2}\right)^{4/3}}, \nonumber \\
    h^{zz}= \lambda^{-2/3}  \frac{\left(1 + \frac{r^2}{4 \omega^2}\right)^{1/3}}{\left(1 - \frac{r^2}{4 \omega^2}\right)^{1/3}}.
    \label{Eq:Principal_directions2}
\end{align}

Let us analyze the properties of this fluid required to simulate the geometry. The first thing we notice is that the velocity of the fluid becomes infinite at $r = r_C$, where it also changes its sign. Hence the fluid system is singular; the $r<r_C$ and $r>r_C$ parts of the system are disconnected. However, the CTCs are living entirely in the exterior region of the metric so it may still appear that the analogue system can locally simulate CTCs. However, looking at the speeds of sound we identify an additional issue. The speeds of sound also diverge at $r=r_C$, but moreover $c_r^2=c^2h^{rr}$ and $c_z^2=c^2 h^{zz}$ become negative for $r>r_C$, so $c_r$ and $c_z$ become purely imaginary. This means that we no longer have wave-like behaviours, or in other words causal signalling, in those directions within the analogue systems. The whole acoustic picture appears to break down. In that sense, the $r=r_C$ cylinder can be understood as a sort of ``domain wall":  it separates the interior region in which we have causal signalling from the region in which we have abnormal (exponentially amplified or attenuated) behaviour of the putative sound-like excitations.

To design a realistic situation, imagine that we start with a fluid at rest and incite a rotation around the $z$-axis with the intention of evolving towards the G\"odel geometry. In order to do so, we need to obtain a configuration in which there is a separation between a clockwise and an anticlockwise rotating part of the fluid, separated by a surface at $r = r_C$ where the velocity needs to approach infinity and moreover the speed of sound also blows up. These requirements essentially imply that the hydrodynamic description of the BEC breaks down. Furthermore, because of the infinite fluid velocity at the $r=r_C$ surface, no signal could cross this surface. However, the physical parameters of the BEC are well defined for $r>r_C$, where one finds CTCs for the BEC excitations. On the one hand, one would need an imaginary sound speed in that region. This can be attained in BECs with attractive interactions~\cite{Wieman2001,Duine2000}. On the other hand, two of the effective anisotropic masses of the BEC must be negative. The peculiarity of a particle with a negative mass is that it accelerates backwards when pushed forward\footnote{Notice that such negative masses are not fundamental, and thus need not result in tachyonic instabilities.}. However, it has been shown experimentally that it is indeed possible to achieve such strange behaviour and create particles with negative effective masses~\cite{Khamehchi2017}.

Therefore, we have a surprising situation. The excitations of a quite strange BEC, with attractive interactions (which in principle would appear to forbid wave phenomena) combined with some negative anisotropic masses, end up behaving as if these excitations live in a perfectly Lorentzian world displaying CTCs. The situation would be equivalent in any other anisotropic fluid, not necessarily quantum\footnote{It is true, however, that engineering a classical fluid to display ``negative mass" excitations might be much more complicated, perhaps even impossible in practice.}. One would just need that $c_r^2$ and $c_z^2$ become negative in some region while $c_\phi^2$ stays positive. Roughly speaking, this ensures that the $r$ and $z$ coordinates acquire the same signature as the $t$ coordinate, leaving the angular coordinate $\phi$ as the coordinate of different signature, i.e. the time coordinate, and CTCs will develop. From the perspective of the internal observers inside the fluid, ``time" would be what for an external (laboratory) observer is simply the angular coordinate. 

\begin{figure}
\begin{center}
\includegraphics[width=0.75 \textwidth]{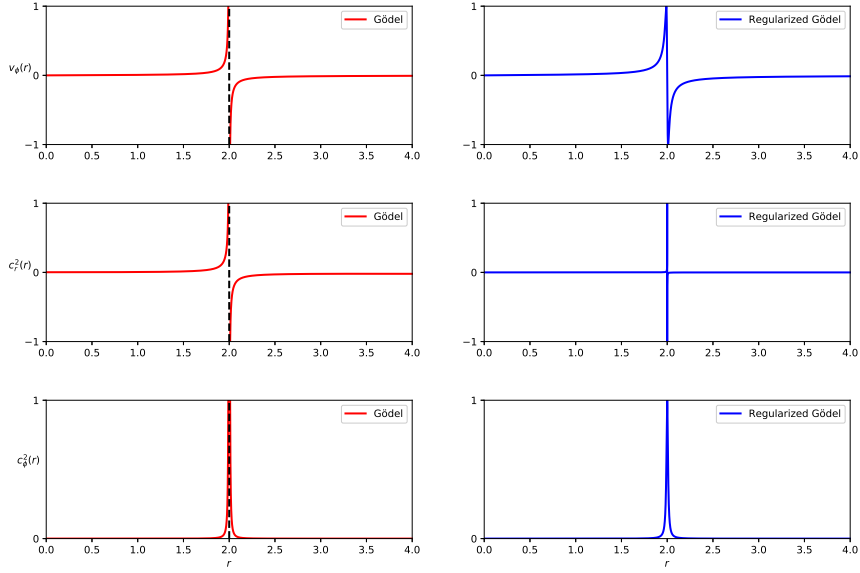}
\caption{The left panel represents the anisotropic speeds of sound and the azimuthal velocity of the fluid required to perform an analogue simulation of G\"odel's geometry for $\omega = 1$. All of them display a vertical asymptote at $r = r_C$. The right panel displays the corresponding anisotropic speeds of sound and the azimuthal velocity of the fluid for the regularized G\"odel geometry introduced in the text, also for $\omega = 1$ and for the regularization parameter $\epsilon = 0.01$. Whereas the graphics on the left panels all blow up at $r=r_C$, the ones in the right panels are smooth everywhere. All functions are normalized to a maximum value of $1$ in the plot.}
\label{Fig:plots-godel}
\end{center}
\end{figure}

Going back to G\"odel's metric, one could be tempted to modify the parameters of the analogue model and regularize the divergences. A simple example would be the following profiles where a suitable regulating parameter $\epsilon \ll 1$ is introduced (for simplicity we restrict our fluid to be effectively two-dimensional):
\begin{align}
    & v_{\phi} = \frac{r}{\sqrt{2} \omega}  \frac{1 - \frac{r^2}{4 \omega^2}}{ \left( 1 - \frac{r^2}{4 \omega^2} \right)^2 + \epsilon^2}, \\
    & c_r^2 = \left( 1 + \frac{r^2}{4 \omega^2}\right)^2 \frac{1 -\frac{r^2}{4 \omega^2}}{\left( 1 -\frac{r^2}{4 \omega^2}\right)^2 + \epsilon^2}, \\
    & c_{\phi}^2 = \frac{1 + \frac{r^2}{4 \omega^2}}{\left(1 - \frac{r^2}{4 \omega^2}\right)^2 + \epsilon^2}.
\end{align}
However, it is straightforward to see that the associated acoustic metric (a cousin of G\"odel's metric) is not a regular Lorentzian metric now, it is degenerate at $r=r_C$. Still, strange as it may seem, this acoustic system does approach G\"odel metric for $r \gg r_C$. These speeds of sound and velocity of the fluid as well as the corresponding ones for pure G\"odel are plotted in Fig.~\ref{Fig:plots-godel}.

A note of caution might be in order at this point. Although this analysis suggests that it is possible to simulate CTCs in an analogue system (a fluid in this case), these CTCs are, in a sense, trivial. From the point of view of the external observer, the creation of these CTCs corresponds simply to declaring that the internal observer is using in the internal system an angular coordinate as time coordinate. From now on, we will refer to this kind of CTCs as trivial, to distinguish them from CTCs that appear in the causal future of a causally well-behaved region, which are the most interesting ones from a physical point of view. Hawking characterized this type of spacetimes geometrically as those with a compactly generated Cauchy horizon~\cite{Hawking1991}\footnote{A compactly generated Cauchy horizon is a Cauchy horizon such that the past extension of its generators enters and remains within a compact subset of the manifold.}. 

Spacetimes with non-trivial CTCs can thus be understood as those with a smooth transition from a region without CTCs to one with CTCs. From an analogue point of view, it seems that the simulation of such non-trivial cases is not possible. Indeed, they would require either a non-regular Lorentzian metric, which could be reproduced within an analogue model, or a well-defined Lorentzian metric but requiring some divergences in the analogue model. The former case would not really constitute a clear proof of principle of the possibility of simulating spacetimes with CTCs, since the CTCs could be understood to be an artifact of the non-smoothness of the metric and thus, in a sense, spurious or at least not directly related to a (non-analogue) relativistic equivalent. On the other hand, the latter case has already been discussed and, based on the example of G\"odel's metric, seems to correspond to trivial CTCs at best, since the two regions need to be causally disconnected.  
   
To finish this section let us consider the possibility of simulating a geometry which approaches G\"odel's metric only in a finite range of the laboratory time $t$. To our knowledge, this geometry does not have a General Relativistic counterpart, in the sense that it is not a solution to the Einstein equations with the energy-momentum tensor of a known matter content. Such behaviour could be achieved with the help of a modulating function $f(t)$, such that we can write a metric
\begin{equation}
    ds^2 = -dt^2 + \frac{dr^2}{1 + f(t) \frac{r^2}{4 \omega^2}} + r^2 \left( 1 - f(t) \frac{r^2}{4 \omega^2}\right) d \phi^2 + dz^2 - f(t) \frac{\sqrt{2}}{\omega} r^2 dt d \phi ,
    \label{Eq:Godel_finite_time}
\end{equation}
where we can choose $f(t)$ to be a function with compact support, for instance
\begin{equation}
    f(t) = \exp \left[ \frac{- \sigma}{(t_B-t) (t-t_A)}\right], 
    \label{Eq:Time_modulating}
\end{equation}
which is non-vanishing for $t \in (t_A,t_B)$. Hence, the metric represents a flat spacetime outside this interval, and develops CTCs within the region $(t_A,t_B)$. In order to confine the CTCs to a compact region of space also, one could force the metric component $g_{\phi \phi}$ to take negative values only within a finite interval of the $r$-component, for example through the following replacement 
\begin{equation}
    g_{\phi \phi} =  r^2 \left( 1 - f(t) \frac{r^2}{4 \omega^2}\right) \longrightarrow  r^2 \left( 1 - f(t) e^{-\frac{r^2}{\sigma^2}}\frac{r^2}{4 \omega^2}\right) ,
\end{equation}
where $\sigma$ must to be sufficiently large in order for the function $1 - e^{-\frac{r^2}{\sigma^2}}\frac{r^2}{4 \omega^2}$ to display two zeros. This new geometry exhibits CTCs  that are confined within a finite region of spacetime. However, for the same arguments explained above, it is not possible to simulate them as acoustic metrics since the fluid would be required to develop a singular velocity. This again illustrates our more general point that it seems impossible to generate metrics with non-trivial CTCs through an analogue metric.

\section{A survey of some spacetimes displaying CTCs amenable to analogue gravity simulation}
\label{Sec:Survey}

In this section we are going to present some geometries containing CTCs which we think are conceptually simpler than G\"odel spacetime. Most of these geometries can be found somehow in the literature.  However, we think that it is worthy to revise them and present them in a unified way, so that they are amenable to be analyzed from the analogue gravity perspective. First, we will start describing the geometry that results from combining two warp drives and displays CTCs. Motivated by the properties of this geometry, we will introduce a family of spacetimes which are simpler but encapsulate their main geometric features. Finally, we will discuss probably the most paradigmatic example of spacetime containing CTCs: the so-called Misner spacetime. This spacetime is used as a proxy to more convoluted analysis, since its chronological horizon is usually considered to have the general properties a chronological horizon has. Although for most practical purposes this is true, we will put special emphasis here on the fact that Misner spacetime is topologically non-trivial as a manifold (otherwise it could not contain CTCs as we will explain). In higher dimensions, (like 3+1 spacetime dimensons) it is possible to have spacetimes with CTCs exhibiting a trivial topology. From an analogue gravity perspective, this makes life much easier for their simulation.

\subsection{CTCs engineering through warp-drive bubbles}
\label{Subsec:CTC-warp drives}

Warp drives were originally introduced by Alcubierre~\cite{Alcubierre1994}. They are based on disturbing a given spacetime within a compact region in such a way that for observers outside that region, observers inside of it move with superluminal speeds. They can be thought as ``tachyonic'' bubbles that propagate faster than light for external observers. The simplest metrics representing warp drives can be written using the N\'atario's line element:
\begin{equation}
    ds^2 = -dt^2 + \delta_{ij} c^{-2} \left( dx^{i} - v^{i} dt\right) \left(dx^{j} - v^{j}  dt\right), 
    \label{Eq:Natario}
\end{equation}
There are some warp drives with non-zero lapse function or with a non-euclidean metric, although we will not focus on them. 

In this way, the metric of a warp drive is nicely adapted to be simulated with the acoustic metric of a BEC, as described above, or acoustic metrics showing up in other fluids. This idea has already been suggested in the literature, see for instance~\cite{Fischer2002,Finazzi2011}. Essentially, we need to identify the velocity of the fluid with the shift functions entering the warp drive element. For concreteness, we can think of a warp-drive bubble whose profile acquires the following form
\begin{equation}
    v^{i} (t, x^i) = \delta^{i}_{\ x} u (t) f \left( \sqrt{ \left(x-x(t)\right)^2 + y^2 + z^2 }\right)
    \label{Eq:Velocity}
\end{equation}
where $f(x)$ is a compact support function, describing the profile of the bubble which is peaked around the points of the trajectory 
\begin{equation}
    x(t) = x(0) + \int^t_0 dt' u(t'), 
    \qquad 
    y=z=0.
\end{equation}
As we just say, the simulation of a single warp drive in a acoustic analogue system is direct~\cite{Barcelo2011}. One just need to generate a spacetime region at which the velocity of the flow exceeds the speed of sound. From the internal perspective this allows to travel from one point to another at velocities higher than that of sound (which remember takes the role of the speed of light). It is convenient to rewrite the warp drive metric in Eq.~\eqref{Eq:Natario} as a perturbation of the flat spacetime metric $\eta_{\mu \nu}$ as
\begin{equation}
    g_{\mu \nu} = \eta_{\mu \nu} + b_{\mu \nu}, 
    \label{Eq:Natario2}
\end{equation}
with $b_{\mu \nu}$ having the following non-vanishing components:
\begin{align}
    & b_{00}  = u^2 (t) f^2\left( \sqrt{ \left(x-x(t)\right)^2 + y^2 + z^2 }\right), \nonumber \\
    & b_{01} = b_{10} = - u (t) f \left( \sqrt{ \left(x-x(t)\right)^2 + y^2 + z^2 }\right).
\end{align}

Now, although a single bubble warp drive by itself does not result in any chronology or causality violations, as already noticed in~\cite{Everett1995} it is relatively easy to engineer a spacetime that contains CTCs by taking advantage of having two warp-drive bubbles in dimensions higher than $1+1$. The idea is similar to the way in which one can send information to the past with a pair of tachyon particles in flat spacetime~\cite{Rolnick1969}. What is it then the clash, if any, between warp drive metrics and their analogue simulations? 

We want to construct regular spacetime geometries based on a combination of two warp drives, in such a way that they contain CTCs. After finding such geometries we will analyze whether it is possible to reproduce them within an analogue model in Sec.~\ref{Sec:Analogue-simulations}. The simplest such construction that one can think of \emph{a priori} involves two warp-drive bubbles in a 1+1 dimensional setting with trivial $\mathbb{R}^2$ topology. In the remain of this subsection, we are going to discuss for a moment this 1+1 potential construction. First, we will naively present it. Then, we will illustrate the obstruction that one finds when one tries to formalize this construction. After that, we will show that this problem cannot be circumvented by presenting some theorems showing that this construction is actually not possible. Finally, we will conclude this Subsection by explaining how this construction can be extended to $D+1$ spacetime dimensions with trivial topology without problems.

\begin{figure}
\begin{center}
\includegraphics[width=0.5 \textwidth]{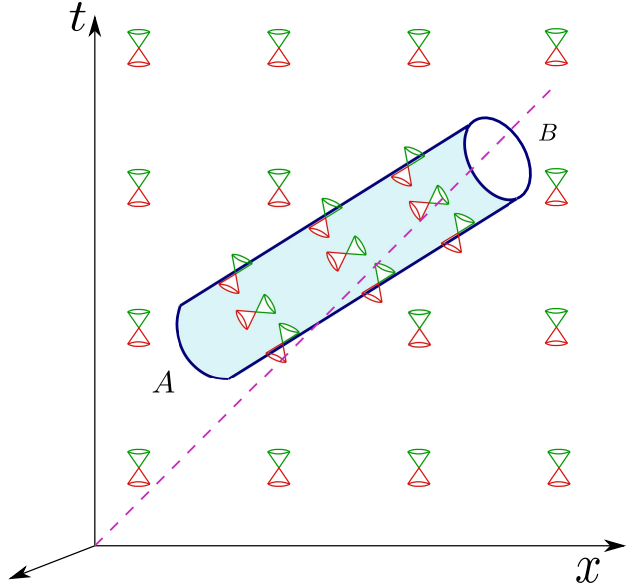}
\caption{The figure represents a warp drive tube that starts in a location A and ends in location B. The underlying spacetime can be considered $D+1$ dimensional with the warp-drive bubble moving in a straight line (along the $x$-coordinate, in the picture). In 1+1 dimensions the causal cone would become just two crossing lines but we keep the cone symbol for clarity. This is the simplest construction of a warp drive and as described in the text, it can be simulated in an analogue gravity model without further problems.}
\label{Fig:warpdrive-forward}
\end{center}
\end{figure}

Let us begin with the most naive way in which one might try to make this configuration. Let us consider a 1+1 dimensional Minkoswski background. Let us choose an inertial reference frame $S$ with Cartesian coordinates $(t,x)$. Furthermore, let us choose two events $A$ and $B$ such that they are spacelike separated, with coordinates $(t_A,x_A)$ and $(t_B,x_B)$, respectively. Without loss of generality, let us assume that $t_B>t_A$. This setup is represented pictorially in Fig.~\ref{Fig:warpdrive-forward}. We can engineer a warp drive tube connecting the two events. Notice that the lightcones inside the tube are modified with respect to the Minkowskian reference. Furthermore, we emphasize that the tube has some thick walls at which the light cones experience a tilting effect. It is precisely on those walls where the stress energy tensor supporting these configurations necessarily develops some energy conditions violations~\cite{Alcubierre1994}. We emphasize that the trajectory as seen from outside the tube appears to be spacelike. Of course, observers going from $A$ to $B$ within the bubble would be following strictly timelike trajectories. 

\begin{figure}
\begin{center}
\includegraphics[width=0.5 \textwidth]{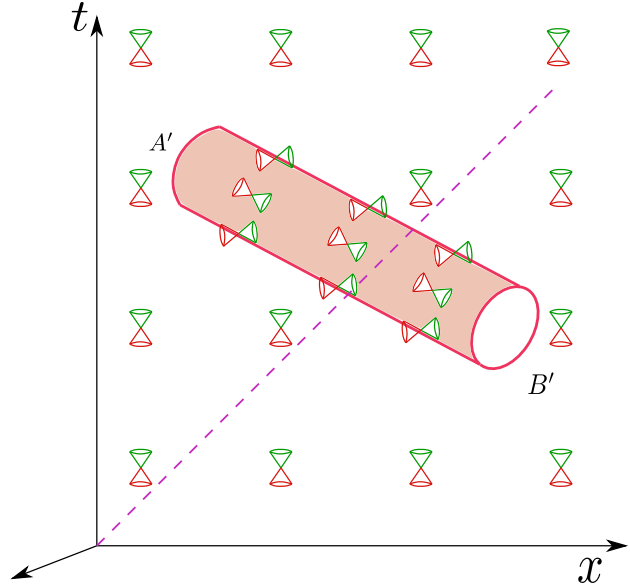}
\caption{The figure represents a warp drive tube that starts in a location A' and ends in location B'. The underlying spacetime can be considered $D+1$ dimensional with the warp-drive bubble moving in a straight line (along the $x$-coordinate, in the picture). In 1+1 dimensions the causal cone would become just two crossing lines but we keep the cone symbol for clarity.}
\label{Fig:warpdrive-backward}
\end{center}
\end{figure}

\begin{figure}
\begin{center}
\includegraphics[width=0.5 \textwidth]{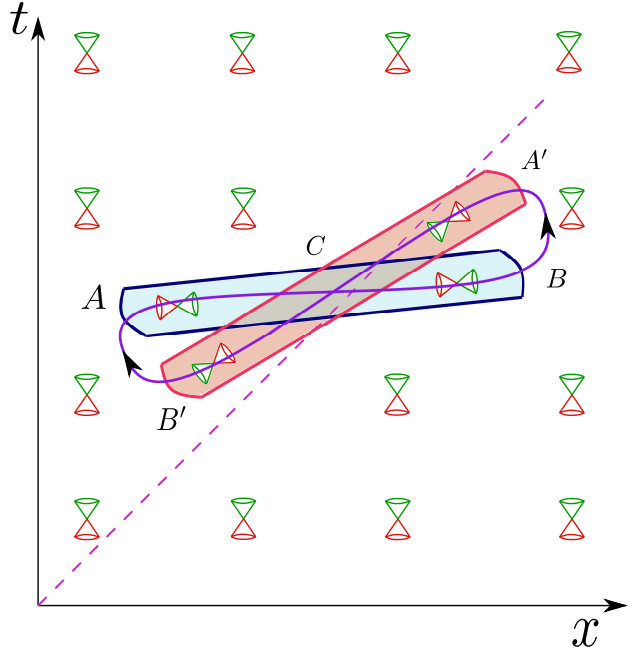}
\caption{We represent here two warp drives in 1+1 dimensions and in such a configuration that they appear to allow for the formation of CTCs. The purple curve represents a generic CTC on this background. The problem with this 1+1 configuration is that the metric in the region where the two warp-drive bubbles cross is ill-defined. As described in the text the simplest geometry with CTCs is either one with a $S^1 \times \mathbb{R}$ topology or with topology $\mathbb{R}^{D+1}$ in $D+1$ dimensions, with $D>1$. In this latter case, we just need to engineer the two warp drives bubbles in different parallel planes to avoid the crossing.}
\label{Fig:warpdrive-crossing}
\end{center}
\end{figure}

If one can construct this warp drive, from a purely general relativistic perspective it is also possible to construct an equivalent warp drive configuration in which the coordinate time goes to the past instead of the future~\cite{Everett1995}. Let us write down such a metric explicitly. Let us start constructing a warp drive metric as the one just described but using another inertial reference frame $S'$ moving with velocity $v$ in the $x$ direction. Let us denote with a prime the cartesian coordinates of the reference frame $S'$, i.e. $(t',x')$. In these coordinates, the metric of the bubble takes the simple form $g'_{\mu \nu}$ of Eq.~\eqref{Eq:Natario}. To find the metric in the coordinates $(t,x)$ adapted to the inertial frame $S$, we simply need to perform a boost of velocity $-v$ in the $x$-axis, with $v$ the relative velocity between $S$ and $S'$. In this way, the resulting metric reads  
\begin{equation}
    g_{\mu \nu} = \eta_{\mu \nu} + c_{\mu \nu}, 
    \label{Eq:backwards_bubble}
\end{equation}
where $c_{\mu \nu}$ is transformed from the prime coordinates to the unprimed ones by an ordinary Lorentz transformation. We highlight that its functional form differs from the $b_{\mu \nu}$ tensor introduced in Eq.~\eqref{Eq:Natario2}. Actually, we can find its functional form by performing a Lorentz boost in the $x$-direction of velocity $-v$, where $v$ is the relative velocity between both frames $S$ and $S'$. Explicitly, it is described by the following linear transformation in : 
\begin{equation}
  \left( \Lambda \right)^{\mu} _{\ \nu}  = \left(
  \begin{array}{cc} 
  \cosh \phi \ &  \sinh \phi  \\  
  \sinh \phi \ & \cosh \phi 
  \end{array}
  \right), \qquad \textrm{with} \qquad  \tanh \phi = v,
\end{equation}
Writing down the transformation we obtain the following $c_{\mu \nu}$ tensor
\begin{align}
    & c_{00} = u^2(t'(t,x),x'(t,x),y-y_0,z) \cosh^2 \phi - 2 u (t'(t,x),x'(t,x),y-y_0,z) \cosh \phi \sinh \phi , \\
    & c_{01} = c_{10} =  u^2(t'(t,x),x'(t,x),y-y_0,z) \cosh \phi \sinh \phi \nonumber \\
    & - u (t'(t,x),x'(t,x),y-y_0,z) \cosh^2 \phi -  u (t'(t,x),x'(t,x),y-y_0,z) \sinh^2 \phi, \\
    & c_{11} = u^2(t'(t,x),x'(t,x),y-y_0,z) \sinh^2 \phi - 2 u^2(t'(t,x),x'(t,x),y-y_0,z) \sinh \phi \cosh \phi.     
\end{align}
Notice that the functions $t'$ and $x'$ depend on the coordinates $t,x$ in a non trivial manner, and we need to rewrite them in terms of such coordinates. In fact, the coordinates $ \{ x^{\mu} \}$ are related to the coordinates $\{ x^{\prime  \mu} \}$ through a Lorentz transformation from $S$ to $S'$ in which the Lorentz matrix is precisely $\Lambda^{\mu}_{\ \nu}$. Explicitly, the change of coordinates reads
\begin{align}
    & t' = t \cosh \phi  - x \sinh \phi, \\
    & x' = t \sinh \phi  + x \cosh \phi .
\end{align}
Writing everything explicitly without fixing a particular trajectory and shape for the bubbles would not be very illustrative, hence we simply keep everything indicated as done above. We emphasize that one just needs to choose a profile for the bubble and the velocities in order to be able to write down explicitly the metric in global coordinates by substituting in the expressions above. In generic terms we have build a warp drive travelling backwards in coordinate time $t$, pictorially represented in Fig.~\ref{Fig:warpdrive-backward}. Note however that one can easily check that the Lorentz transformation we have applied make the new warp drive metric to take a different form from Natario's line element. For the arguments that follow the precise form of the tubes will not be relevant. 
Now setting up a combination of a ``forward" and a ``backward" warp-drive bubbles one can attempt to build a time machine. Let us explicitly illustrate this. We can first set up a ``forward" warp drive allowing a faster-than-light travel from A to B. Once the traveller has exit the bubble at B he could immediately enter in a new warp drive, now of a ``backward'' type,
and travel from $A'$ to $B'$. Using another spacelike trajectory, as seen from the external Minkowski spacetime, this second warp drive can take the time-traveller to an event $B'$ in pass of the initial event $A$. In this way a CTC is completed. This setup is pictorially represented in Fig.~\ref{Fig:warpdrive-crossing}. 

\begin{figure}
\begin{center}
\includegraphics[width=0.3\textwidth]{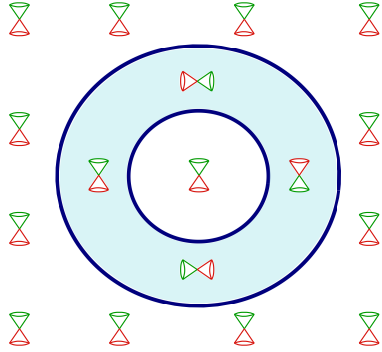}
\caption{We represent here the setup described in the text that already shows the difficulty present when trying to build CTCs in a spacetime with a trivial topology. The shaded region represents the region of abnormal behaviour of the lightcones. It seems impossible to regularize the lightcones without removing points from the spacetime, otherwise the metric would need to vanish at some point and hence it would not be a regular spacetime. }
\label{Fig:Toy_example}
\end{center}
\end{figure}

However, there is a problem concerning this construction. There is a region, the crossing region $C$, at which one would need to have two different metrics. Actually, this translates into a singular point where the metric is not defined. It is natural then to pose the following question: is it possible to disentangle the crossing point moving around the starting and ending points of the bubbles or/and playing with their shapes and the specific forms of their velocities $v_1(t),v_2(t)$, in such a way that we find a completely regular Lorentzian metric containing CTCs? The answer to this question is negative. To understand why, it is useful to consider a toy geometry which nicely illustrates the obstruction. Imagine that we write down a geometry which is that of flat spacetime at every point except for a circle of radius one around the origin in a given set of Cartesian coordinates $(t,x)$. At such circle, we make the lightcones to make an angle of $45$ degrees with the circle at every point. Clearly such geometry is singular since there is a jump in the metric. Even if we try to smooth such geometry by giving the circle a finite size and converting it into a disk, we can find a curve along which the lightcones make a rotation of $360$ degrees and, hence, it is impossible to have a smooth metric in the region enclosed by such curve or regularize it in some way. This is depicted schematically in Fig.~\ref{Fig:Toy_example}.

Now we are in position of stating the following theorem:

\emph{Theorem:} Let $(\mathcal{M},g)$ be a two dimensional simply connected spacetime (being $\mathcal{M}$ the smooth manifold and $g$ its metric). Then, the causality condition automatically holds. 

Recall~\cite{Hawking1973} that a spacetime is said to satisfy the \emph{chronological condition} if it does not contain any closed timelike curves, and it is said to obey the \emph{causality condition} if there are no closed non-spacelike curves. The idea of the proof is already contained in the observation that we have made above: having the structure of lightcones enclosing a compact region, it is impossible to push them inwards or outwards that region without making them zero or singular. The formalization of this statement can be found in~\cite{Oneill1983}. 

It is possible to even prove a stronger result. In Lorentzian geometry there exist a hierarchy of causality conditions where each of them is stronger than the previous ones. Although the chronology condition is the weakest of such conditions, followed by the causality condition, and they are enough to rule out closed non-spacelike curves, one can still think of spacetimes that are ``arbitrarily'' close to containing closed causal curves. Hence, these set of stronger conditions attempts to formalize these notion of ``almost having closed curves''~\cite{Wald1984,Hawking1973}. 

The \emph{strong causality condition}~\cite{Wald1984}, which is obeyed by a spacetime if for every point $p$ and every neighbourhood $N$ of $p$, there exists a neighbourhood $O$ contained in $N$ such that no causal curve intersects $O$ more than once. By essentially the same arguments exhibited in our proof, one can strengthen the result and prove that every two dimensional time-orientable simply connected spacetime is strongly causal (see Lemma 14.34 of~\cite{Oneill1983}). 

Actually, it is even possible to strengthen this result under the same hypothesis. A spacetime is said to be \emph{stably causal} if there exists a timelike vector field $t^\mu$ such that the Lorentz metric defined as $\Tilde{g}_{\mu \nu} = g_{\mu \nu} - t_{\mu} t_{\nu}$ (which has larger lightcones than $g_{\mu \nu}$ at every point) contains no closed causal curves. It can be proved that a spacetime is stably causal if and only if it admits a globally defined time-function, i.e. a function that is strictly increasing along each future directed causal curve~\cite{Hawking1973}. It is also possible to prove that stable causality implies strong causality, i.e. it is a stronger condition. Hence, stable causality is a stronger condition and one might wonder whether it is possible to prove that every two dimensional simply connected spacetime obeys it without further assumptions. In~\cite{Beem1996} the affirmative answer is provided in the form of Theorem 3.43, where it is shown that every simply connected two dimensional spacetime $(\mathcal{M},g)$ is indeed stably causal. Two dimensional spacetimes display further special properties, see for instance footnote 5 of~\cite{SanchezCaja2021}.

This concludes our discussion of $1+1$ dimensional spacetimes and warp drives, where we have shown that it is impossible to build CTCs while preserving the $\mathbb{R}^2$ topology. The statement that no two-dimensional topologically trivial time-machine can be built is in agreement with the literature. All the spacetimes that are pathological from a causal point of view in $1+1$ spacetime dimensions have a non-trivial topology. For instance, Misner's spacetime~\cite{Misner1967} (see Subsec.~\ref{Subsec:Misner}), which is used as a proxy to generic spacetimes containing a chronological horizons, has the topology of a cylinder $\mathbb{R} \times \mathbb{S}^1$. 

In $D+1$ dimensions the previous obstruction does not longer apply. We can perfectly avoid the warp drive crossing by translating one of the tubes in an orthogonal direction: i.e. the forward and backward warp drives can be set up to live in different parallel planes. A complete metric containing CTCs based on warp drives can be designed by just translating in a transverse direction one of the two warp drives.

\subsection{Generalized warp-drive regions}
\label{Subsec:Extended-Warpdrives}

The metric just described is somewhat convoluted. We construct now a family of geometries encoding the main properties of these spacetimes in a much simpler way. These geometries are not built to be a solution to the Einstein equations with a given energy-momentum tensor. However, apart from being useful for our discussion, they are suitable for further analysis concerning aspects like semiclassical effects on chronologically ill-behaved spacetimes. Let us detail this construction. 

The idea is to consider the manifold $\mathbb{R}^{3} \times \mathcal{N}$, with $\mathcal{N}$ a given manifold that we can think of as either being compact or non-compact. Let us choose polar coordinates $(t,r,\phi)$ for $\mathbb{R}^3$ and another suitable set of coordinates $\{ x^{n} \}$ parametrizing $\mathcal{N}$. For the sake of simplicity, we will choose the metric to factorize into the Lorentzian geometric structure of $\mathbb{R}^{3} $ and a given Riemannian metric $g_{mn}$ on $\mathcal{N}$. Although realistic analogue gravity constructions will reproduce only 2+1 or 3+1 configurations, we will keep the discussion as general as possible. Let us write down the following metric 
\begin{equation}
ds^2 = - dt^2 + dr^2 - 2 r^2 f(r)g(t) dt d\phi + r^2 \left( 1 - f(r) g(t) \right)  d\phi^2 + g_{mn} dx^m dx^n,
\end{equation}
with $f(r)$ and $g(t)$ functions with the properties that we now discuss. The function $f(r)$ needs to be such that $f(0) = f(\infty) = 0$. However, we need it to change the  given compact region $ r \in (r_a,r_b)$. This resembles the discussion on G\"odel's spacetime but now with CTCs strictly confined to a finite range of the radial coordinate. There are many different choices of $f(r)$ that do the job. For instance, a relatively simple choice is that of a bump function
\begin{equation}
      f(r) = \left\{
\begin{array}{ll}
     0 & r \notin (r_a,r_b) \\
      \exp \left[ \frac{-\sigma^2}{(r_b-r) (r-r_a)} \right] & r \in (r_a,r_b)\\
\end{array} 
\right. 
\end{equation}
For any $\sigma$ the function is identically $0$ outside the interval $(r_a,r_b)$, and becomes positive inside it. Actually, we can relax the compact support condition and choose a broader class of functions that allow an analytical treatment. For example, we can think of a P\"oschl-Teller like function
\begin{equation}
f(r) = \sigma^2 \sech^2 \left( r^* - r_0\right), \qquad  \textrm{with} \qquad r^*= r + r_0\ln(r/r_0).
\end{equation}

Now we have come to the point of discussing the function $g(t)$. Depending on whether we want to confine CTCs to a given finite range of the $t$-coordinate, have a flat non-compact region free of such CTCs and then glue it to another region containing CTCs or have a whole cylinder in spacetime in which the $\phi$ coordinate acts as a time, we must choose different profiles for $g(t)$. Let us sketch here the three cases. 

\begin{figure}
\begin{center}
\includegraphics[width=0.3 \textwidth]{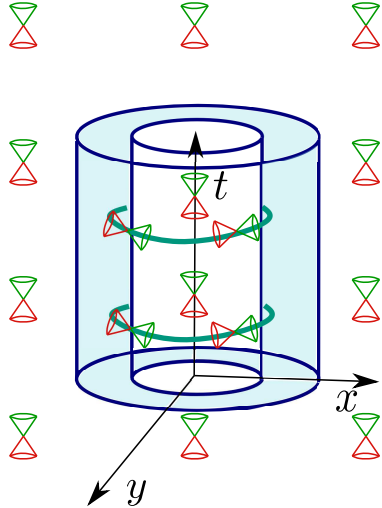}
\caption{We represent here the geometries described in the text. The different cases, corresponding essentially to the finite, the semi-infinite and the infinite cylinder display this same structure, with the difference relying on whether the cylinder is confined to lie within a compact interval $[t_a,t_b]$, a semi-infinite interval $[t_C, \infty)$ or the whole real line $(-\infty, \infty)$. The grey-shaded region lying within the inner and outer surfaces of the cylinder is the region where the angular coordinate $\phi$ becomes timelike, since the lightcones tilt enough for orbits of $\phi$ to be timelike curves. Outside the shaded region this abnormal behaviour of the ligthcones disappears and they are simply (or approximately as we have discussed in the text) flat spacetime lightcones. It must be noticed the difference between this higher dimensional case and the two-dimensional case discussed above: the existence of an additional dimension (the vertical $t$-axis in the figure) allows to regularize the lightcones within the core of the cylinder without needing to change the topology.}
\label{Fig:Cylinder}
\end{center}
\end{figure}

\begin{enumerate}
    \item \textbf{CTCs through every} $ \bf{t=\textrm{\textbf{constant}}}$ \textbf{surface:} In this case, we can simply choose to have $g(t) = 1$ with any of the $f(r)$ choices mentioned above and we will have that $\phi$ is timelike for a whole cylinder of spacetime $ I \times \mathbb{R} \times S^1$. A pictorical representation of these geometries can be found in Fig.~\ref{Fig:Cylinder}.
    
    \item \textbf{Gluing a flat region to a semi-infinite cylinder containing CTCs:} In this case, we are interesting in choosing a function $g(t)$ that vanishes at $t< t_c$ and grows up to one for $t \rightarrow \infty $. Again, we can relax the condition of vanishing for $t<t_c$ by a condition of taking an almost zero value. A suitable choice of such functions is the following. For a function such that it exactly vanishes for $t< t_c$ we can simply choose a bump function as the ones we have chosen before 
    \begin{equation}
          g(t) = \left\{
    \begin{array}{ll}
          0 & t < t_c \\
           \exp \left[ \frac{-\sigma^2}{t-t_c} \right] & t \in (t_c,\infty) \\
    \end{array} 
    \right. 
    \end{equation}
    If we do not insist on having an identically vanishing function we can replace the bump function by a hyperbolic tangent or an inverse tangent, i.e. 
    \begin{align}
        & g(t) = \frac{1}{2} \left( 1 + \tanh \left( \frac{t}{\sigma} \right) \right), \\
        & g(t) = \frac{1}{\pi} \left( \frac{\pi}{2} + \tan^{-1} \left( \frac{t}{\sigma} \right) \right),
    \end{align}
    where $\sigma$ modulates how fast do this function reach $1$, i.e., how fast the region where $\phi$ becomes timelike is reached by growing in the $t$ coordinate. 
    
    \item \textbf{CTCs within a compact region:} In this case, we want to choose the function $g(t)$ that is $0$ outside a compact interval $[t_a,t_b]$ and approaches $1$ in a compact region. Again, we might choose it to be not exactly zero for the purpose of having analytic expression. This choice of functions is completely parallel to the discussion of $g(r)$ functions above and hence we just write down again the functions for the sake of completeness. On the one hand, if we insist on having zero outside the compact time interval we might choose a bump function of the form 
    \begin{equation}
          g(t) = \left\{
    \begin{array}{ll}
          0 & t \notin (t_a,t_b) \\
          \exp \left[ -\frac{\sigma^2}{(t_b-t) (t-t_a)} \right] & t \in (t_a,t_b).\\
    \end{array} 
    \right. 
    \end{equation}
    On the other hand, if we allow to have an almost zero function outside $(t_a,t_b)$ instead of an exact zero (which we again emphasize that is enough for the geometry to have the qualitative properties that we have discussed) then we can simply use a Gaussian or a P\"oschl-Teller like function: 
    \begin{align}
    & g(t) =  \exp \left[ - \frac{(t-t_0)^2}{2 \sigma^2}\right], \\
    & g(t) = \sigma^2 \sech^2 \left( t - t_0\right),
    \end{align}
where we have introduced the center of the region at which $\phi$ becomes timelike defined as $t_0 = (t_b-t_a) /2 + t_a$. 
\end{enumerate}

This concludes our catalogue of toy-geometries containing CTCs which are inspired on those generated by the warp-drive tubes described in Subsec.~\ref{Subsec:CTC-warp drives}.

\subsection{Misner and Misner-like spacetimes}
\label{Subsec:Misner}

\begin{figure}[H]
\begin{center}
\includegraphics[width=0.25\textwidth, height=0.45\textwidth]{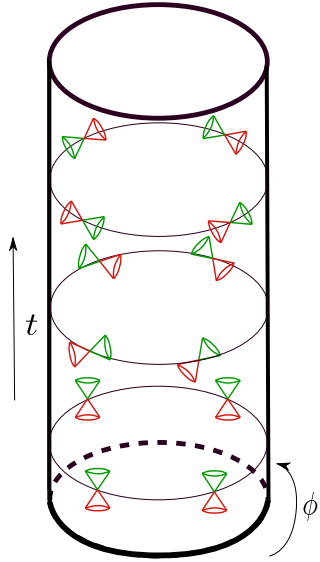}
\caption{ We represent here a cylinder $\mathbb{R} \times \mathbb{S}^1$ endowed with the light cone structure of Misner spacetime. At $t= - \infty$, the lightcones seem to have a ``normal" causality in the sense of not allowing closed causal curves. They get tilted as we move from $t=-\infty$ to $t=0$, where one of the generators of the light cones is perpendicular to $t$. Hence, at $t=0$, we got the first closed causal curve: a light ray that is confined to such circle. This $t=0$ slice corresponds to the circle in the middle of the three drawn. For $t \geq 0$, we enter into the regime of abnormal causal behaviour containing closed causal curves. Notice that, although we are still using conic symbols for the light cones, we are again in 1+1 spacetime dimensions and they are not strictly ``cones''.} 
\label{Fig:misner_spacetime}
\end{center}
\end{figure}

The obstruction to build spacetimes with CTCs in a $1+1$-dimensional spacetime that we found in Subsec.~\ref{Subsec:CTC-warp drives} was topological in nature. If we relax the condition of having a trivial $\mathbb{R}^2$-topology, we can build CTCs in spacetimes of such dimensionality. Actually, the archetypal example of spacetime with a chronological horizon (compactly generated Cauchy horizon) is precisely the so-called Misner spacetime~\cite{Misner1967,Hawking1973}. It has a $\mathbb{R} \times \mathbb{S}^1$ topology. This spacetime is taken generically as a proxy to study chronological horizons. Given coordinates $(t,\phi)$ on the cylinder, the metric of this spacetime can be written as 
\begin{equation}
    ds^2 = -2 dt d \phi - t d \phi^2
\end{equation}
Misner spacetime can be regarded as a simpler lower dimensional version of the Taub-NUT spacetime~\cite{Hawking1973}. This spacetime is such that for $t<0$ it displays a normal causal behaviour. However, at $t=0$ we find a compactly generated Cauchy horizon separating this chronologically ``safe" region from the chronologically ``sick'' region $t \geq 0$ displaying closed causal curves. A pictorical representation of this spacetime can be found in Fig.~\ref{Fig:misner_spacetime}. 
Imagine now that instead of endowing the cylinder with the Misner metric we endow it with the flat metric. A well-posed problem in this spacetime could be given by simply considering data on a constant $t= t_0$ slice which are $2 \pi$ periodic in $\phi$. After evolving forward and backwards such data, we would end up with a global solution. However, we should now recall that what we call ``time'' and what we call ``space'' in a $1+1$ dimensional spacetime is a choice. Pictorially, it corresponds to declaring which is the direction of time flow.

\begin{figure}[H]
\begin{center}
\includegraphics[width=0.29 \textwidth]{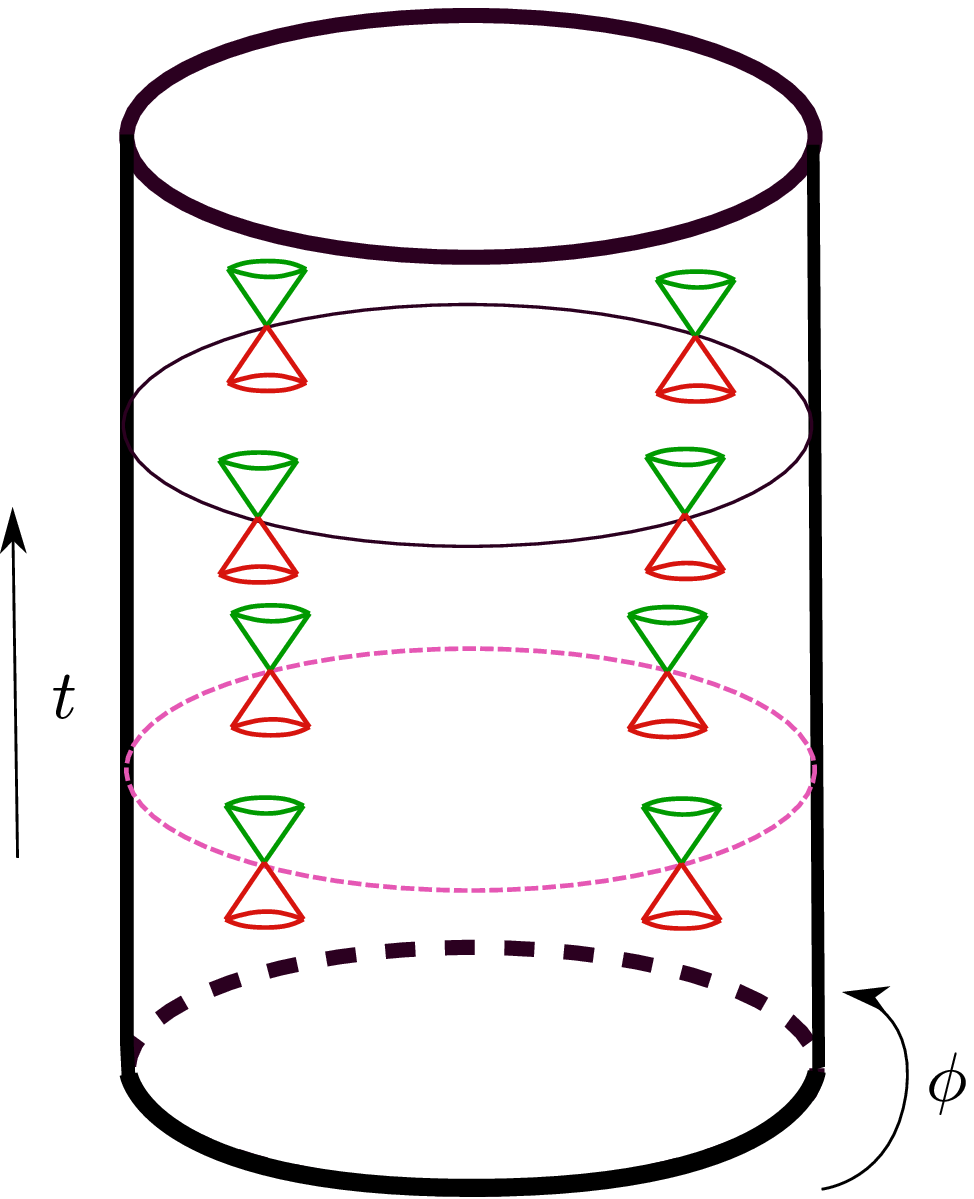}
\includegraphics[width=0.3 \textwidth]{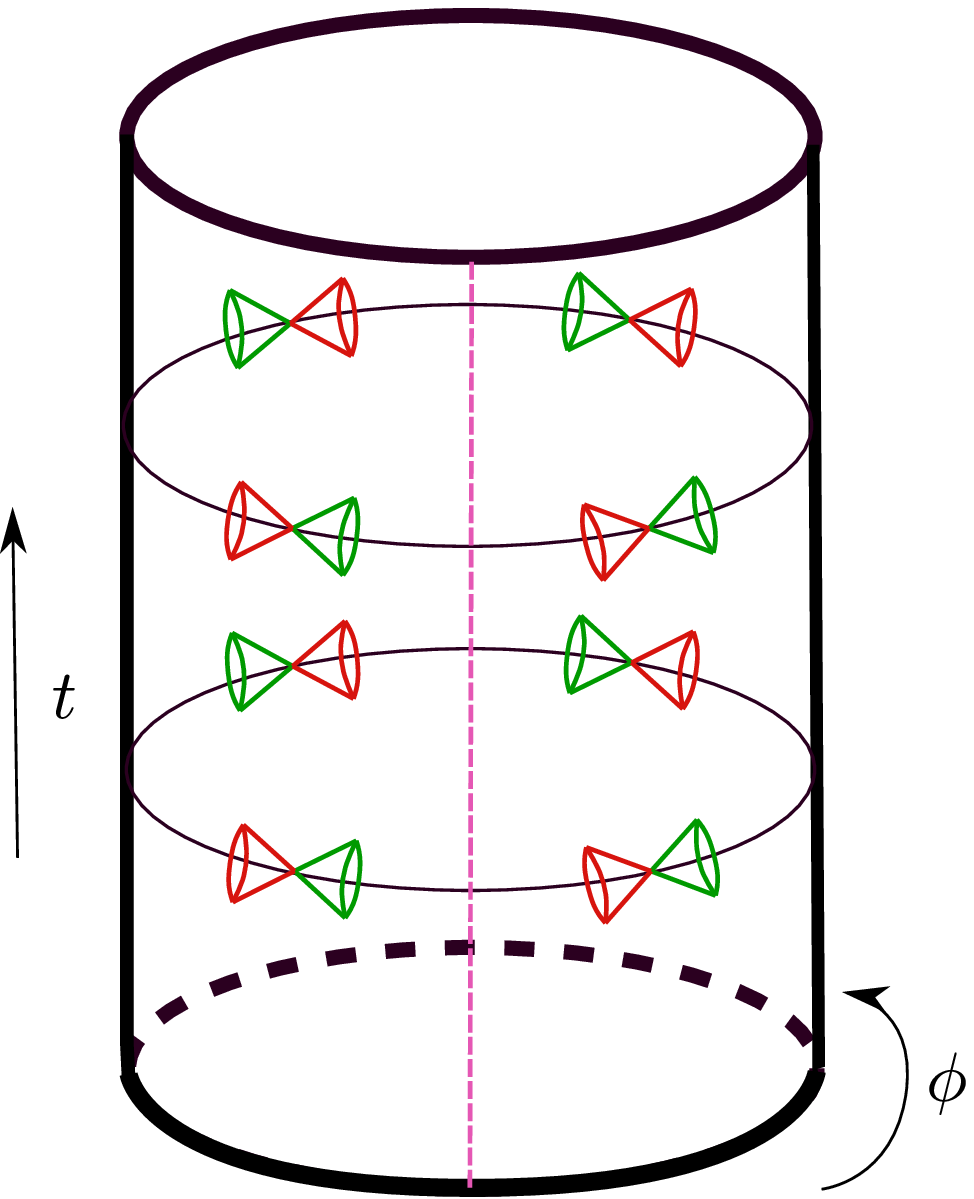}
\caption{The cylinder endowed with a flat metric allows two alternative perspectives regarding the causal description of the spacetime: a spacelike ring evolving in time or as a line configuration evolving in time periodically. The two alternatives are differentiated just by the choice of what we call time and what we call space. We have marked on dotted pink line the putative surfaces on which we would put initial data according to our choice of time and space directions.}
\label{Fig:flat_cylinder}
\end{center}
\end{figure}

In this second perspective, in which $\phi$ plays the role of time, a Cauchy data given on a constant $\phi$-line would require to fullfill additional ``self-consistency'' constraints to have a well-defined evolution~\cite{Novikov1989b,Echeverria1991,Friedman1990b}. However, we know that such data exist in this metric because we know that global solutions exist and we have simply interchanged our choice of time and space! All this discussion serves for the purpose of illustrating that this configuration can be interpreted either as a spacelike ring evolving in time or as a line configuration evolving in a cyclic time. This is depicted in Fig.~\ref{Fig:flat_cylinder}

This example serves us to illustrate that, among the potential geometries displaying CTCs that one might consider, there are some of them which are mild in a sense. Whereas Misner spacetime is undoubtedly pathological since it develops CTCs from a chronologically well-behaved region, the second example of a cylinder with flat metric and CTCs all along is trivial since, as we have explained, they correspond to a ``weird" choice of time direction. These CTCs are trivial in the sense we have described above. As we will explain, whereas these ones are amenable to simulation, the Misner-like CTCs are not.

\section{Analogue gravity simulations: Attempts to simulate CTCs}
\label{Sec:Analogue-simulations}

Now that we have presented different Lorentzian geometries with CTCs based on G\"odel's spacetime, connecting warp-drive bubbles and generalizations of them, and Misner and Misner-like spacetimes, we can directly discuss whether these geometries are amenable to be simulated in BECs and other acoustic systems. 

First of all, it is interesting to realize that the flat cylinder with time $\phi$ periodically identified $\phi \sim \phi + 2 \pi$ can perfectly be simulated in an analogue system, for example with a fluid inside an effectively 1-dimensional ring and declaring that the physical angular coordinate is a time coordinate. This very simple configuration serves to illustrate that, contrary to the standard lore, one can actually simulate spacetimes with CTCs in analogue systems. The same trick could be done in principle in $2+1$ and $3+1$ configurations although it would be slightly more complicated. The choice of time as an angular coordinate $\phi$ requires that the standard time and the additional spatial coordinates must acquire the same signature in the analogue model. In this way, the effective causality in the analogue system can have CTCs. 

However, as emphasized above, this kind of ``eternal" CTCs as seen from an external observer is trivial in a sense. They correspond to declaring that an angular coordinate for the external observer is being used as a time coordinate for the internal observer. The price to pay is needing to impose extra ``self-consistency" conditions to obtain a well-posed problem. However, in the cases of eternal CTCs, those self-consistency constraints can be satisfied trivially by some global solutions. This can be easily seen by noticing that, if we choose the non-periodic time coordinate $t$, global solutions exist as the Cauchy development of some well-posed initial value problem on surfaces of constant $t$. Those global solutions automatically verify this self-consistency constraint after a trivial relabelling of the time coordinate. 

Thus, the most interesting case consists in simulating geometries with a chronological horizon: those exhibiting a causally well-behaved region separated from a causally pathological region by a compactly generated Cauchy horizon. Here we point out that an acoustic realization of those configurations always involves the divergence of the velocity of the fluid. All the types of geometries described in Sec.~\ref{Sec:Survey} involve tilting the sound cones until one is able to move infinitely rapidly with respect to the laboratory reference frame (and actually, even more). But this is forbidden by the presence of the actual speed of light limit. Furthermore, to travel backwards in laboratory time would be forbidden even in a Galilean world (formally the $c \rightarrow \infty$ limit of a Lorentzian world). This means that, although for an internal observer the configurations illustrated in Figs.~\ref{Fig:warpdrive-forward} and~\ref{Fig:warpdrive-backward} are on equal footing, they are actually not due to the background structure that an external observer has access to. Hence, whereas the metric structure in Fig.~\ref{Fig:warpdrive-forward} can be simulated, the metric structure from Fig.~\ref{Fig:warpdrive-backward} cannot. 

This situation seems to be generic in configurations in which a region with CTCs is separated from a region that does not have CTCs: the generation of CTCs is always accompanied by a divergence in the fluid velocity. In addition there will be changes of sign in several anisotropic speeds $c^2$'s. As the G\"odel example from Sec.~\ref{Sec:Godel} illustrates nicely, these changes can be achieved through a divergence in the analogue system parameters or passing through a zero value. Exactly the same happens for the geometries explored in Sec.~\ref{Sec:Analogue-simulations}: it is a generic feature of the attempts to simulate this kind of geometries. In the regular case, in which the sounds of speed pass through zero, the analogue system can be perfectly defined but it is the acoustic metric that becomes a non-regular Lorentzian metric and thus, from a relativistic point of view, describes a completely different situation.

We argue that the obstacle that we have identified in simulating spacetimes with chronological horizons applies to generic analogue systems and not only to the concrete acoustic systems that we have been analyzing. The argument is based on the following property which holds, to the best of our knowledge, for all analogue systems explored in the literature so far. Independently of the specific form of the analogue physical metric\footnote{Here by ``physical'' we mean the analogue metric controlling the relevant fields, as opposed to an eikonal metric controling just geometrical optics behaviour.} in a particular analogue system, one can always take an eikonal approximation. The eikonal dispersion relation can always be written using a normalized inverse metric (one that obeys $g_{\textrm{(E)}}^{tt}=-1$), resulting in
\begin{equation}
    g^{\mu \nu}_{\textrm{(E)}} p_{\mu} p_{\nu} = E^2 - 2 g^{ti}p_i + g^{i j} p_{i} p_{j},
\end{equation}
where $p_{\mu} = \left(E, \boldsymbol{p} \right)$ and $\boldsymbol{p} = p^{i}$. Now, the central point is that the remaining components of the eikonal inverse metric are always identified with some properties of the substratum system~\cite{Barcelo2011}: in the fluid we have considered, they can be identified with the velocities of sound and the fluid; in dielectric media they are identified with the permittivity tensor $\epsilon_{ij}$, \emph{etc}. 

We expect something similar also to hold for Weyl-semimetals which seem to be good analogues~\cite{Volovik2016}. In such case, the identification of the divergent quantities with physical parameters is subtle, since one identifies the metric-components with some vectors characterizing the structure of the interacting Weyl-point~\cite{Volovik2017}. We leave the detailed analysis on the divergences that appear in Weyl-fermions analogues of chronological horizons to future work. This subtle point in the identification of physical properties is what leads to the impossibility of simulating geometries containing CTCs in the eikonal approximation, such as the G\"odel geometry studied in~\cite{Fiorini2021}. The analysis in~\cite{Sabin2016,Sabin2018} seem to deal with what we have called trivial CTCs. As such, their conclusions that these spacetimes can be simulated seem to be consistent with our results here.

We assume that such identification is always true for the analogues of interest. We can now finish the proof under this assumption. The spacetimes with chronological horizons that we have been analyzing are such that one can always introduce a chart of coordinates in which an angular coordinate $\phi$, in the causally well-behaved region, becomes a time coordinate in the causally pathological region. On the boundary, which constitutes the chronological horizon, it becomes null. In order to analyze what this implies for the eikonal metric components in these coordinates, let us write the metric and its inverse in matrix form. For the argument that follows it is sufficient to restrict ourselves to the $(t,\phi)$ coordinate sector (with $t$ the time coordinate of the well-behaved region). Then we have:
\begin{align}
    g = \left(
  \begin{array}{cc} 
  g_{tt} \ &  g_{t \phi}  \\  
  g_{t \phi} \ & g_{\phi \phi}
  \end{array}
  \right), \qquad \left( g\right)^{-1} = \frac{1}{g_{tt} g_{\phi \phi} - \left( g_{t \phi} \right)^2} 
  \left( \begin{array}{cc} 
  g_{\phi \phi} \ &  - g_{ t \phi}  \\  
  - g_{t \phi} \ & g_{t t}
  \end{array}
  \right).
\end{align}
The fact that $\phi$ becomes null at the chronological horizon means that $g_{\phi \phi}$ vanishes there. But this in turn implies that $g^{tt}$ becomes zero. To obtain the normalized eikonal metric with $g_{\textrm{(E)}}^{tt}=-1$ from the inverse physical metric, one needs to divide the inverse metric by a conformal factor $\Omega$ which must also tend to zero in order to keep the division normalized. But then, being divided by this same zero, the other components of the eikonal metric will blow up at the chronological horizon : $g_{\textrm{(E)}}^{ti}= g^{ti}/\Omega$, $g_{\textrm{(E)}}^{ii}=g^{ii}/\Omega \to \infty$. As these components are directly identified with physical properties of the system, the presence of this type of horizons would necessarily require a singular configuration from the substratum perspective. Even though the general relativistic metric components, and so the hypothetical physical analogue metric components, can be well defined, divergences in the components of the inverse eikonal metric forbid the construction of a proper analogue model, i.e. one with a non-singular substratum. The detailed physical properties that might diverge depend on the specific characteristics of the analogue system. In fact, it would be interesting to perform detailed analyses of how concrete analogue systems break down when trying to engineer them until they are ``on the verge" of forming a chronological horizon. 

This impossibility of constructing analogue models of a chronological horizon is resonant with Hawking's Chronology Protection proposal~\cite{Hawking1991}. In the next section, we will discuss the interplay of this conjecture with our analysis. 

\section{Chronology protection: lessons from analogue gravity}
\label{Sec:Chronology-Protection}

In General Relativity, all available evidence suggests that the construction of geometries with superluminal behaviour\footnote{Defining superluminality in curved backgrounds is not straightforward. We will simply mean signals that propagate faster than the speed of light in flat spacetime.} requires violations of energy conditions~\cite{Penrose1993,Olum1998,Visser1998,Visser1999,Gao2000}. For instance, the construction of warp drives such as the ones described above requires exotic matter~\cite{Alcubierre1994}. Now, let us assume for a moment that exotic matter exists or could be created and manipulated at will for this purpose. The arguments that follow will be kinematic in character and not depend on dynamical considerations. Under the previous hypothesis we could in principle engineer a geometry containing a warp drive. The standard perspective provided by classical General Relativity is that travelling faster than light entails the possibility of travelling to the past. For instance, the construction of a forward-in-coordinate-time warp drive is equivalent to the construction of a backward-in-coordinate-time warp drive due to the diffeomorphism invariance of General Relativity. Under all these assumptions and tricks, it would be almost trivially possible to build a time machine.

Although this reasoning is of course logically correct, our analogue gravity analysis shows that even under the assumption of freely available and manipulable exotic matter, internal observers could face obstructions when trying to engineer CTCs in the presence of a fundamental background structure non-observable to them in standard non-extremal situations. From their perspective, the inability to engineer such time machines could manifest itself differently depending on the specific ``high-energy" theory, for example as instabilities, singular behaviours or phase transitions. In all cases, this would constitute a dramatic breakdown of their internal description of physics in terms of low-energy effective fields, such as the metric description itself, which would become useless in these regimes. For instance, the construction of the forward and backward warp drives would seem equivalent to these internal engineers. However, from the laboratory perspective, they are not equivalent at all, and in fact it is easy to see that the backward warp drive would be impossible to design. To sum up, travelling at speeds greater than the speed of sound is perfectly possible, but one cannot send signals backwards in time. 

Interestingly, one can also reverse the logic of the previous paragraph. When internal observers try to construct superluminal configurations, at the same time they implicitly probe whether different inertial observers are indeed equivalent, or said in other words: whether there is a more fundamental causality underneath the General Relativity description. It might happen that some inertial observers can construct a superluminal warp drive while others encounter a mysterious effect obstructing an equivalent construction. This non-equivalence would point at a breakdown of basic relativistic principles at this level of description. Our analysis also illustrates that beating the speed of light would not necessarily imply the possibility of travelling backwards in time and constructing a time machine. One might find insurmountable troubles in the way, caused by some fundamental substratum. We will come back to this point below.

From this perspective, it is interesting to note that hints of causality protection already exist in General Relativity at the classical level, and even stronger hints at the semiclassical level. These hints led Hawking to conjecture his famous Chronology Protection Conjecture~\cite{Hawking1991}. Essentially, Hawking conjectured that the fundamental laws of physics would forbid time travel, even though from the point of view of General Relativity it is an open possibility. After some counter-analysis by Kim and Thorne~\cite{Kim1991}, it appears that the consensus reached by the community is that, within the framework of classical and semiclassical gravity, one cannot settle the question of whether the formation of a region of spacetime developing CTCs from a non-pathological initial-data region is forbidden or not. In that sense, although the majority of the community seems to expect that these spacetimes are forbidden, they believe that the ultimate reason for it lies deep in the quantum gravity regime. See~\cite{Visser2002} for a comprehensive review.

All in all, it appears that the consensus reached by the community is that, within the framework of classical and semiclassical gravity, one cannot settle the question of whether the formation of a region of spacetime developing CTCs from a non-pathological initial-data region is forbidden or not. In that sense, although the majority of the community seems to expect that these spacetimes are forbidden, they believe that the ultimate reason for it lies deep in the quantum gravity regime (see~\cite{Visser2002} for a comprehensive review). 

Let us now discuss a specific contribution of this paper to the Chronology Protection notion that goes beyond the analogue systems examined explicitly above. Although these analogue systems obey dynamics by no means similar to the dynamics of general relativity, one can conceive a system that does have a general-relativistic-like dynamics. This is what we would call Emergent Gravity: gravity emerging as a collective excitation of microscopic degrees of freedom living in a deeper layer of reality with a Galilean or Minkowskian structure~\cite{Barcelo2021}. Because of this underlying structure, these theories would contain a direct mechanism forbidding the formation of a chronological horizon. This would arguably provide the simplest Chronology Protection mechanism beyond classical or semiclassical General Relativity. Actually, let us stress that this idea does not rely on a ``quantum gravity" argument. 

We recall two observations: i) Following Rosen~\cite{Rosen1940a,Rosen1940b}, General Relativity itself can be formulated alternatively as a non-linear theory of gravitons over a flat Minkowski background; ii) Weyl-Tranverse (WT) gravity~\cite{Alvarez2006} is a theory indistinguishable from General Relativity, at least classically, which needs the existence of an externally fixed volume-form. In both theories, the Minkowski metric or the volume-form do not acquire any operational role. At the present moment, there are no observational arguments to prefer standard General Relativity over the less conventional Rosen or WT-gravity perspectives. However, the non-evidence of exotic matter and causally strange behaviours could be taken as evidences of a high-energy Chronology Protection mechanism. From this point of view, the existence of a background structure in Rosen and WT-gravity offer frameworks which incorporate a Chronology Protection mechanism in a rather natural way.

It is also worth emphasizing the parallel between our conclusions and the ones found in~\cite{Emparan2021a} on how the AdS/CFT correspondence seems to enforce chronology protection. 
One can understand the AdS/CFT model as an emergent gravity theory, in which the microscopic theory is a CFT living in a causally well-behaved spacetime, and leads to the emergence of gravity in the bulk. The effective geometric description when attempting to engineer CTCs consists of two causally disconnected regions in the bulk, thus making any such CTCs harmless for the causally-well behaved region. This is completely parallel to what we have found in our analogue gravity discussion of G\"odel's spacetime, for instance.

Hence, it seems that the mechanism by which Chronology Protection is implemented in emergent approaches to gravity has some degree of universality or independence from the microscopic details of the models, at least, as long as such models display a well-behaved causal structure, corresponding to the broad type I category of Carlip's classification of emergent models of gravity~\cite{Carlip2013}.

One final lesson that can be learnt from the examples presented here, as already briefly mentioned earlier, is that superluminal signal propagation does not automatically lead to the possibility of building time machines or, more generally, abnormal causal behaviour. There are claims in the literature that any kind of superluminality is problematic, see for instance~\cite{Dolgov1998,Adams2006,Camanho2014}. The opposite view was defended, for example, by~\cite{Liberati2001,Babichev2007}. Geroch has argued nicely for the position that superluminality does not automatically imply causal troubles~\cite{Geroch2010}, by making the point that chronologically pathological spacetimes cannot correspond to the evolution of a well-posed initial value problem. From our point of view, the key point is that superluminality is not sufficient for causal pathologies to appear. The theory must allow for further manipulations. In the presence of a background structure, these manipulations depend on the properties of the background structure and could thus well be forbidden, for example in the rather general case in which this background structure has a causal structure (Galilean or Minkowskian) on its own.

\section{Summary and conclusions}
\label{Sec:Summary}

This work started by inquiring whether it is possible to simulate geometries with CTCs within the analogue gravity program. We began analyzing probably the most famous geometry possessing CTCs: the G\"odel spacetime. There are claims in the literature that it is impossible to reproduce spacetimes with CTCs of any sort. The reason is that the analogue system must be embedded in a laboratory, and the laboratory is living in a (locally) Minkowskian spacetime which has a well-behaved causality. This causality must be inherited by the analogue model. 

We have seen that this argument needs further qualifications. We have shown that certain configurations with CTCs can be perfectly reproduced in an analogue system, and in fact G\"odel's spacetime belongs to this category. However, this class corresponds in a specific sense to trivial CTCs. The really interesting situations correspond to spacetimes with CTCs which moreover also have chronological horizons. Such horizons separate the region with CTCs from the region without CTCs. It is this category of spacetimes that lead to pathologies when we attempt to simulate them. 

We have presented a simple catalogue of analytical spacetime metrics containing CTCs that we have later attempted to simulate in analogue gravity setups. In all the examples analyzed, the presence of a chronological horizon leads to an insurmountable obstacle for its implementation in analogue gravity: one or more physical parameters of the analogue system must diverge at the chronological horizon. In some cases, these divergences can be smoothed out, but this destroys the regularity of the Lorentzian metric description which leads to the interpretation of the physical system as a gravitational analogue in the first place. We have focused on an acoustic system in which it is essentially the velocity of the fluid which must diverge in order to create the required tilting of the sound cones. We have shown that it is perfectly possible to simulate geometries allowing superluminal behaviours such as warp drives. However, this does not imply directly that one can build an analogue time machine, as we have discussed in detail. It is in fact the formation of a chronological horizon which is forbidden in the analogue implementation, since it is not possible to create a warp drive travelling backward in laboratory time.  

The obstructions found by exploring the analogue gravity implementation of CTCs resonate with Hawking's idea of a Chronology Protection mechanism in semiclassical General Relativity. From the point of view presented here, such protection mechanisms arise naturally in frameworks for Emergent Gravity. In these emergent frameworks, there exists a background structure with a more fundamental underlying causality, which naturally prevents the type of manipulations required to create chronological pathologies.

Finally, we sum up three important lessons from the present work. i) Superluminality itself does not imply the possibility of abnormal causal behaviour such as time travel; ii) Problems in the analogue implementation of chronological horizons appear due to the relative tilting of the causal cones; iii) The current observational absence of chronological pathologies in our universe is naturally explained in frameworks in which there is a fixed underlying causality beyond the local General Relativistic modifications explored so far.

\acknowledgments{The authors thank Grisha Volovik for useful conversations and for sharing unpublished notes on the simulation of G\"odel spacetime. The authors thank Carlos Sab\'{i}n and Franco Fiorini for helpful correspondence. GGM and CB thank Luis Garay, Miguel S\'anchez and Valent\'in Boyanov for very useful conversations. GGM thanks Roberto Empar\'an for an enlightening conversation. Financial support was provided by the Spanish Government through the projects PID2020-118159GB-C43 and PID2020-118159GB-C44, and by the Junta de Andaluc\'{\i}a through the project FQM219. C.B. and G.G.M. acknowledge financial support from the State Agency for Research of the Spanish MCIU through the ``Center of Excellence Severo Ochoa'' award to the Instituto de Astrof\'{\i}sica de Andaluc\'{\i}a (SEV-2017-0709). GGM is funded by the Spanish Government fellowship FPU20/01684.}

\newpage

\bibliography{chronology_stuff_analogues}

\begin{thebibliography}{58}%
\makeatletter
\providecommand \@ifxundefined [1]{%
 \@ifx{#1\undefined}
}%
\providecommand \@ifnum [1]{%
 \ifnum #1\expandafter \@firstoftwo
 \else \expandafter \@secondoftwo
 \fi
}%
\providecommand \@ifx [1]{%
 \ifx #1\expandafter \@firstoftwo
 \else \expandafter \@secondoftwo
 \fi
}%
\providecommand \natexlab [1]{#1}%
\providecommand \enquote  [1]{``#1''}%
\providecommand \bibnamefont  [1]{#1}%
\providecommand \bibfnamefont [1]{#1}%
\providecommand \citenamefont [1]{#1}%
\providecommand \href@noop [0]{\@secondoftwo}%
\providecommand \href [0]{\begingroup \@sanitize@url \@href}%
\providecommand \@href[1]{\@@startlink{#1}\@@href}%
\providecommand \@@href[1]{\endgroup#1\@@endlink}%
\providecommand \@sanitize@url [0]{\catcode `\\12\catcode `\$12\catcode
  `\&12\catcode `\#12\catcode `\^12\catcode `\_12\catcode `\%12\relax}%
\providecommand \@@startlink[1]{}%
\providecommand \@@endlink[0]{}%
\providecommand \url  [0]{\begingroup\@sanitize@url \@url }%
\providecommand \@url [1]{\endgroup\@href {#1}{\urlprefix }}%
\providecommand \urlprefix  [0]{URL }%
\providecommand \Eprint [0]{\href }%
\providecommand \doibase [0]{http://dx.doi.org/}%
\providecommand \selectlanguage [0]{\@gobble}%
\providecommand \bibinfo  [0]{\@secondoftwo}%
\providecommand \bibfield  [0]{\@secondoftwo}%
\providecommand \translation [1]{[#1]}%
\providecommand \BibitemOpen [0]{}%
\providecommand \bibitemStop [0]{}%
\providecommand \bibitemNoStop [0]{.\EOS\space}%
\providecommand \EOS [0]{\spacefactor3000\relax}%
\providecommand \BibitemShut  [1]{\csname bibitem#1\endcsname}%
\let\auto@bib@innerbib\@empty
\bibitem [{\citenamefont {Barcel{\'o}}\ \emph {et~al.}(2011)\citenamefont
  {Barcel{\'o}}, \citenamefont {Liberati},\ and\ \citenamefont
  {Visser}}]{Barcelo2011}%
  \BibitemOpen
  \bibfield  {author} {\bibinfo {author} {\bibfnamefont {C.}~\bibnamefont
  {Barcel{\'o}}}, \bibinfo {author} {\bibfnamefont {S.}~\bibnamefont
  {Liberati}}, \ and\ \bibinfo {author} {\bibfnamefont {M.}~\bibnamefont
  {Visser}},\ }\href {\doibase 10.12942/lrr-2011-3} {\bibfield  {journal}
  {\bibinfo  {journal} {Living Reviews in Relativity}\ }\textbf {\bibinfo
  {volume} {14}},\ \bibinfo {pages} {159} (\bibinfo {year} {2011})}\BibitemShut
  {NoStop}%
\bibitem [{\citenamefont {Volovik}(2009)}]{Volovik2009}%
  \BibitemOpen
  \bibfield  {author} {\bibinfo {author} {\bibfnamefont {G.}~\bibnamefont
  {Volovik}},\ }\href {https://books.google.es/books?id=6uj76kFJOHEC} {\emph
  {\bibinfo {title} {The Universe in a Helium Droplet}}},\ International Series
  of Monographs on Physics\ (\bibinfo  {publisher} {OUP Oxford},\ \bibinfo
  {year} {2009})\BibitemShut {NoStop}%
\bibitem [{\citenamefont {Kolobov}\ \emph {et~al.}(2021)\citenamefont
  {Kolobov}, \citenamefont {Golubkov}, \citenamefont {Mu\~noz~de Nova},\ and\
  \citenamefont {Steinhauer}}]{Steinhauer2021}%
  \BibitemOpen
  \bibfield  {author} {\bibinfo {author} {\bibfnamefont {V.~I.}\ \bibnamefont
  {Kolobov}}, \bibinfo {author} {\bibfnamefont {K.}~\bibnamefont {Golubkov}},
  \bibinfo {author} {\bibfnamefont {J.~R.}\ \bibnamefont {Mu\~noz~de Nova}}, \
  and\ \bibinfo {author} {\bibfnamefont {J.}~\bibnamefont {Steinhauer}},\
  }\href {\doibase 10.1038/s41567-020-01076-0} {\bibfield  {journal} {\bibinfo
  {journal} {Nature Phys.}\ }\textbf {\bibinfo {volume} {17}},\ \bibinfo
  {pages} {362} (\bibinfo {year} {2021})}\BibitemShut {NoStop}%
\bibitem [{\citenamefont {Garay}\ \emph {et~al.}(2000)\citenamefont {Garay},
  \citenamefont {Anglin}, \citenamefont {Cirac},\ and\ \citenamefont
  {Zoller}}]{Garay1999}%
  \BibitemOpen
  \bibfield  {author} {\bibinfo {author} {\bibfnamefont {L.~J.}\ \bibnamefont
  {Garay}}, \bibinfo {author} {\bibfnamefont {J.~R.}\ \bibnamefont {Anglin}},
  \bibinfo {author} {\bibfnamefont {J.~I.}\ \bibnamefont {Cirac}}, \ and\
  \bibinfo {author} {\bibfnamefont {P.}~\bibnamefont {Zoller}},\ }\href
  {\doibase 10.1103/PhysRevLett.85.4643} {\bibfield  {journal} {\bibinfo
  {journal} {Phys. Rev. Lett.}\ }\textbf {\bibinfo {volume} {85}},\ \bibinfo
  {pages} {4643} (\bibinfo {year} {2000})},\ \Eprint
  {http://arxiv.org/abs/gr-qc/0002015} {arXiv:gr-qc/0002015} \BibitemShut
  {NoStop}%
\bibitem [{\citenamefont {Jannes}\ and\ \citenamefont
  {Volovik}(2015)}]{Jannes2015}%
  \BibitemOpen
  \bibfield  {author} {\bibinfo {author} {\bibfnamefont {G.}~\bibnamefont
  {Jannes}}\ and\ \bibinfo {author} {\bibfnamefont {G.~E.}\ \bibnamefont
  {Volovik}},\ }\href {\doibase 10.1134/S0021364015140052} {\bibfield
  {journal} {\bibinfo  {journal} {JETP Lett.}\ }\textbf {\bibinfo {volume}
  {102}},\ \bibinfo {pages} {73} (\bibinfo {year} {2015})},\ \Eprint
  {http://arxiv.org/abs/1506.00882} {arXiv:1506.00882 [gr-qc]} \BibitemShut
  {NoStop}%
\bibitem [{\citenamefont {Torres}\ \emph {et~al.}(2017)\citenamefont {Torres},
  \citenamefont {Patrick}, \citenamefont {Coutant}, \citenamefont {Richartz},
  \citenamefont {Tedford},\ and\ \citenamefont
  {Weinfurtner}}]{Weinfurtner2016}%
  \BibitemOpen
  \bibfield  {author} {\bibinfo {author} {\bibfnamefont {T.}~\bibnamefont
  {Torres}}, \bibinfo {author} {\bibfnamefont {S.}~\bibnamefont {Patrick}},
  \bibinfo {author} {\bibfnamefont {A.}~\bibnamefont {Coutant}}, \bibinfo
  {author} {\bibfnamefont {M.}~\bibnamefont {Richartz}}, \bibinfo {author}
  {\bibfnamefont {E.~W.}\ \bibnamefont {Tedford}}, \ and\ \bibinfo {author}
  {\bibfnamefont {S.}~\bibnamefont {Weinfurtner}},\ }\href {\doibase
  10.1038/nphys4151} {\bibfield  {journal} {\bibinfo  {journal} {Nature Phys.}\
  }\textbf {\bibinfo {volume} {13}},\ \bibinfo {pages} {833} (\bibinfo {year}
  {2017})},\ \Eprint {http://arxiv.org/abs/1612.06180} {arXiv:1612.06180
  [gr-qc]} \BibitemShut {NoStop}%
\bibitem [{\citenamefont {Braidotti}\ \emph {et~al.}(2021)\citenamefont
  {Braidotti}, \citenamefont {Prizia}, \citenamefont {Maitland}, \citenamefont
  {Marino}, \citenamefont {Prain}, \citenamefont {Starshynov}, \citenamefont
  {Westerberg}, \citenamefont {Wright},\ and\ \citenamefont
  {Faccio}}]{Faccio2021}%
  \BibitemOpen
  \bibfield  {author} {\bibinfo {author} {\bibfnamefont {M.~C.}\ \bibnamefont
  {Braidotti}}, \bibinfo {author} {\bibfnamefont {R.}~\bibnamefont {Prizia}},
  \bibinfo {author} {\bibfnamefont {C.}~\bibnamefont {Maitland}}, \bibinfo
  {author} {\bibfnamefont {F.}~\bibnamefont {Marino}}, \bibinfo {author}
  {\bibfnamefont {A.}~\bibnamefont {Prain}}, \bibinfo {author} {\bibfnamefont
  {I.}~\bibnamefont {Starshynov}}, \bibinfo {author} {\bibfnamefont
  {N.}~\bibnamefont {Westerberg}}, \bibinfo {author} {\bibfnamefont {E.~M.}\
  \bibnamefont {Wright}}, \ and\ \bibinfo {author} {\bibfnamefont
  {D.}~\bibnamefont {Faccio}},\ }\href@noop {} {\enquote {\bibinfo {title}
  {Measurement of penrose superradiance in a photon superfluid},}\ } (\bibinfo
  {year} {2021}),\ \Eprint {http://arxiv.org/abs/2109.02307} {arXiv:2109.02307
  [physics.optics]} \BibitemShut {NoStop}%
\bibitem [{\citenamefont {Barcelo}\ \emph {et~al.}(2003)\citenamefont
  {Barcelo}, \citenamefont {Liberati},\ and\ \citenamefont
  {Visser}}]{Barcelo2003}%
  \BibitemOpen
  \bibfield  {author} {\bibinfo {author} {\bibfnamefont {C.}~\bibnamefont
  {Barcelo}}, \bibinfo {author} {\bibfnamefont {S.}~\bibnamefont {Liberati}}, \
  and\ \bibinfo {author} {\bibfnamefont {M.}~\bibnamefont {Visser}},\ }\href
  {\doibase 10.1142/S0218271803004092} {\bibfield  {journal} {\bibinfo
  {journal} {Int. J. Mod. Phys. D}\ }\textbf {\bibinfo {volume} {12}},\
  \bibinfo {pages} {1641} (\bibinfo {year} {2003})},\ \Eprint
  {http://arxiv.org/abs/gr-qc/0305061} {arXiv:gr-qc/0305061} \BibitemShut
  {NoStop}%
\bibitem [{\citenamefont {Steinhauer}\ \emph {et~al.}(2021)\citenamefont
  {Steinhauer}, \citenamefont {Abuzarli}, \citenamefont {Aladjidi},
  \citenamefont {Bienaim\'e}, \citenamefont {Piekarski}, \citenamefont {Liu},
  \citenamefont {Giacobino}, \citenamefont {Bramati},\ and\ \citenamefont
  {Glorieux}}]{Steinhauer2021b}%
  \BibitemOpen
  \bibfield  {author} {\bibinfo {author} {\bibfnamefont {J.}~\bibnamefont
  {Steinhauer}}, \bibinfo {author} {\bibfnamefont {M.}~\bibnamefont
  {Abuzarli}}, \bibinfo {author} {\bibfnamefont {T.}~\bibnamefont {Aladjidi}},
  \bibinfo {author} {\bibfnamefont {T.}~\bibnamefont {Bienaim\'e}}, \bibinfo
  {author} {\bibfnamefont {C.}~\bibnamefont {Piekarski}}, \bibinfo {author}
  {\bibfnamefont {W.}~\bibnamefont {Liu}}, \bibinfo {author} {\bibfnamefont
  {E.}~\bibnamefont {Giacobino}}, \bibinfo {author} {\bibfnamefont
  {A.}~\bibnamefont {Bramati}}, \ and\ \bibinfo {author} {\bibfnamefont
  {Q.}~\bibnamefont {Glorieux}},\ }\href@noop {} {\enquote {\bibinfo {title}
  {{Analogue cosmological particle creation in an ultracold quantum fluid of
  light}},}\ } (\bibinfo {year} {2021}),\ \Eprint
  {http://arxiv.org/abs/2102.08279} {arXiv:2102.08279 [cond-mat.quant-gas]}
  \BibitemShut {NoStop}%
\bibitem [{\citenamefont {Barcelo}\ and\ \citenamefont
  {Campos}(2003)}]{Barcelo2002}%
  \BibitemOpen
  \bibfield  {author} {\bibinfo {author} {\bibfnamefont {C.}~\bibnamefont
  {Barcelo}}\ and\ \bibinfo {author} {\bibfnamefont {A.}~\bibnamefont
  {Campos}},\ }\href {\doibase 10.1016/S0370-2693(03)00646-4} {\bibfield
  {journal} {\bibinfo  {journal} {Phys. Lett. B}\ }\textbf {\bibinfo {volume}
  {563}},\ \bibinfo {pages} {217} (\bibinfo {year} {2003})},\ \Eprint
  {http://arxiv.org/abs/hep-th/0206217} {arXiv:hep-th/0206217} \BibitemShut
  {NoStop}%
\bibitem [{\citenamefont {Fischer}\ and\ \citenamefont
  {Visser}(2003)}]{Fischer2002}%
  \BibitemOpen
  \bibfield  {author} {\bibinfo {author} {\bibfnamefont {U.~R.}\ \bibnamefont
  {Fischer}}\ and\ \bibinfo {author} {\bibfnamefont {M.}~\bibnamefont
  {Visser}},\ }\href {\doibase 10.1209/epl/i2003-00103-6} {\bibfield  {journal}
  {\bibinfo  {journal} {Europhys. Lett.}\ }\textbf {\bibinfo {volume} {62}},\
  \bibinfo {pages} {1} (\bibinfo {year} {2003})},\ \Eprint
  {http://arxiv.org/abs/gr-qc/0211029} {arXiv:gr-qc/0211029} \BibitemShut
  {NoStop}%
\bibitem [{\citenamefont {Finazzi}(2011)}]{Finazzi2011}%
  \BibitemOpen
  \bibfield  {author} {\bibinfo {author} {\bibfnamefont {S.}~\bibnamefont
  {Finazzi}},\ }\emph {\bibinfo {title} {{Analogue gravitational phenomena in
  Bose-Einstein condensates}}},\ \href@noop {} {Ph.D. thesis},\ \bibinfo
  {school} {SISSA, Trieste} (\bibinfo {year} {2011}),\ \Eprint
  {http://arxiv.org/abs/1208.4729} {arXiv:1208.4729 [gr-qc]} \BibitemShut
  {NoStop}%
\bibitem [{\citenamefont {Everett}(1996)}]{Everett1995}%
  \BibitemOpen
  \bibfield  {author} {\bibinfo {author} {\bibfnamefont {A.~E.}\ \bibnamefont
  {Everett}},\ }\href {\doibase 10.1103/PhysRevD.53.7365} {\bibfield  {journal}
  {\bibinfo  {journal} {Phys. Rev. D}\ }\textbf {\bibinfo {volume} {53}},\
  \bibinfo {pages} {7365} (\bibinfo {year} {1996})}\BibitemShut {NoStop}%
\bibitem [{\citenamefont {Visser}(1998)}]{Visser1997b}%
  \BibitemOpen
  \bibfield  {author} {\bibinfo {author} {\bibfnamefont {M.}~\bibnamefont
  {Visser}},\ }\href {\doibase 10.1088/0264-9381/15/6/024} {\bibfield
  {journal} {\bibinfo  {journal} {Class. Quant. Grav.}\ }\textbf {\bibinfo
  {volume} {15}},\ \bibinfo {pages} {1767} (\bibinfo {year} {1998})},\ \Eprint
  {http://arxiv.org/abs/gr-qc/9712010} {arXiv:gr-qc/9712010} \BibitemShut
  {NoStop}%
\bibitem [{\citenamefont {Hawking}\ and\ \citenamefont
  {Ellis}(2011)}]{Hawking1973}%
  \BibitemOpen
  \bibfield  {author} {\bibinfo {author} {\bibfnamefont {S.~W.}\ \bibnamefont
  {Hawking}}\ and\ \bibinfo {author} {\bibfnamefont {G.~F.~R.}\ \bibnamefont
  {Ellis}},\ }\href {\doibase 10.1017/CBO9780511524646} {\emph {\bibinfo
  {title} {{The Large Scale Structure of Space-Time}}}},\ Cambridge Monographs
  on Mathematical Physics\ (\bibinfo  {publisher} {Cambridge University
  Press},\ \bibinfo {year} {2011})\BibitemShut {NoStop}%
\bibitem [{\citenamefont {Barcelo}\ \emph {et~al.}(2001)\citenamefont
  {Barcelo}, \citenamefont {Liberati},\ and\ \citenamefont
  {Visser}}]{Barcelo2001}%
  \BibitemOpen
  \bibfield  {author} {\bibinfo {author} {\bibfnamefont {C.}~\bibnamefont
  {Barcelo}}, \bibinfo {author} {\bibfnamefont {S.}~\bibnamefont {Liberati}}, \
  and\ \bibinfo {author} {\bibfnamefont {M.}~\bibnamefont {Visser}},\ }\href
  {\doibase 10.1088/0264-9381/18/6/312} {\bibfield  {journal} {\bibinfo
  {journal} {Class. Quant. Grav.}\ }\textbf {\bibinfo {volume} {18}},\ \bibinfo
  {pages} {1137} (\bibinfo {year} {2001})},\ \Eprint
  {http://arxiv.org/abs/gr-qc/0011026} {arXiv:gr-qc/0011026} \BibitemShut
  {NoStop}%
\bibitem [{\citenamefont {Wald}(1984)}]{Wald1984}%
  \BibitemOpen
  \bibfield  {author} {\bibinfo {author} {\bibfnamefont {R.~M.}\ \bibnamefont
  {Wald}},\ }\href {\doibase 10.7208/chicago/9780226870373.001.0001} {\emph
  {\bibinfo {title} {{General Relativity}}}}\ (\bibinfo  {publisher} {Chicago
  Univ. Pr.},\ \bibinfo {address} {Chicago, USA},\ \bibinfo {year}
  {1984})\BibitemShut {NoStop}%
\bibitem [{\citenamefont {Misner}\ \emph {et~al.}(1973)\citenamefont {Misner},
  \citenamefont {Thorne},\ and\ \citenamefont {Wheeler}}]{Misner1974}%
  \BibitemOpen
  \bibfield  {author} {\bibinfo {author} {\bibfnamefont {C.~W.}\ \bibnamefont
  {Misner}}, \bibinfo {author} {\bibfnamefont {K.~S.}\ \bibnamefont {Thorne}},
  \ and\ \bibinfo {author} {\bibfnamefont {J.~A.}\ \bibnamefont {Wheeler}},\
  }\href@noop {} {\emph {\bibinfo {title} {{Gravitation}}}}\ (\bibinfo
  {publisher} {W. H. Freeman},\ \bibinfo {address} {San Francisco},\ \bibinfo
  {year} {1973})\BibitemShut {NoStop}%
\bibitem [{\citenamefont {Godel}(1949)}]{Godel1949}%
  \BibitemOpen
  \bibfield  {author} {\bibinfo {author} {\bibfnamefont {K.}~\bibnamefont
  {Godel}},\ }\href {\doibase 10.1103/RevModPhys.21.447} {\bibfield  {journal}
  {\bibinfo  {journal} {Rev. Mod. Phys.}\ }\textbf {\bibinfo {volume} {21}},\
  \bibinfo {pages} {447} (\bibinfo {year} {1949})}\BibitemShut {NoStop}%
\bibitem [{\citenamefont {Jannes}\ and\ \citenamefont {Volovik}()}]{Volovik}%
  \BibitemOpen
  \bibfield  {author} {\bibinfo {author} {\bibfnamefont {G.}~\bibnamefont
  {Jannes}}\ and\ \bibinfo {author} {\bibfnamefont {G.~E.}\ \bibnamefont
  {Volovik}},\ }\href@noop {} {\enquote {\bibinfo {title} {{G\"odel’s
  universe and time-like curves in condensed matter}},}\ }\bibinfo
  {howpublished} {unpublished}\BibitemShut {NoStop}%
\bibitem [{\citenamefont {Fiorini}\ \emph {et~al.}(2021)\citenamefont
  {Fiorini}, \citenamefont {Hernandez},\ and\ \citenamefont
  {Losada}}]{Fiorini2021}%
  \BibitemOpen
  \bibfield  {author} {\bibinfo {author} {\bibfnamefont {F.}~\bibnamefont
  {Fiorini}}, \bibinfo {author} {\bibfnamefont {S.~M.}\ \bibnamefont
  {Hernandez}}, \ and\ \bibinfo {author} {\bibfnamefont {E.~L.}\ \bibnamefont
  {Losada}},\ }\href {\doibase 10.1103/PhysRevD.104.124009} {\bibfield
  {journal} {\bibinfo  {journal} {Phys. Rev. D}\ }\textbf {\bibinfo {volume}
  {104}},\ \bibinfo {pages} {124009} (\bibinfo {year} {2021})},\ \Eprint
  {http://arxiv.org/abs/2108.12049} {arXiv:2108.12049 [gr-qc]} \BibitemShut
  {NoStop}%
\bibitem [{\citenamefont {Kajari}\ \emph {et~al.}(2004)\citenamefont {Kajari},
  \citenamefont {Walser}, \citenamefont {Schleich},\ and\ \citenamefont
  {Delgado}}]{Kajari2004}%
  \BibitemOpen
  \bibfield  {author} {\bibinfo {author} {\bibfnamefont {E.}~\bibnamefont
  {Kajari}}, \bibinfo {author} {\bibfnamefont {R.}~\bibnamefont {Walser}},
  \bibinfo {author} {\bibfnamefont {W.~P.}\ \bibnamefont {Schleich}}, \ and\
  \bibinfo {author} {\bibfnamefont {A.}~\bibnamefont {Delgado}},\ }\href
  {\doibase 10.1023/B:GERG.0000046184.03333.9f} {\bibfield  {journal} {\bibinfo
   {journal} {Gen. Rel. Grav.}\ }\textbf {\bibinfo {volume} {36}},\ \bibinfo
  {pages} {2289} (\bibinfo {year} {2004})},\ \Eprint
  {http://arxiv.org/abs/gr-qc/0404032} {arXiv:gr-qc/0404032} \BibitemShut
  {NoStop}%
\bibitem [{\citenamefont {Donley}\ \emph {et~al.}(2001)\citenamefont {Donley},
  \citenamefont {Claussen}, \citenamefont {Cornish}, \citenamefont {Roberts},
  \citenamefont {Cornell},\ and\ \citenamefont {Wieman}}]{Wieman2001}%
  \BibitemOpen
  \bibfield  {author} {\bibinfo {author} {\bibfnamefont {E.~A.}\ \bibnamefont
  {Donley}}, \bibinfo {author} {\bibfnamefont {N.~R.}\ \bibnamefont
  {Claussen}}, \bibinfo {author} {\bibfnamefont {S.~L.}\ \bibnamefont
  {Cornish}}, \bibinfo {author} {\bibfnamefont {J.~L.}\ \bibnamefont
  {Roberts}}, \bibinfo {author} {\bibfnamefont {E.~A.}\ \bibnamefont
  {Cornell}}, \ and\ \bibinfo {author} {\bibfnamefont {C.~E.}\ \bibnamefont
  {Wieman}},\ }\href {\doibase 10.1038/35085500} {\bibfield  {journal}
  {\bibinfo  {journal} {Nature}\ }\textbf {\bibinfo {volume} {412}},\ \bibinfo
  {pages} {295–299} (\bibinfo {year} {2001})},\ \Eprint
  {http://arxiv.org/abs/cond-mat/0105019} {arXiv:cond-mat/0105019} \BibitemShut
  {NoStop}%
\bibitem [{\citenamefont {Duine}\ and\ \citenamefont
  {Stoof}(2001)}]{Duine2000}%
  \BibitemOpen
  \bibfield  {author} {\bibinfo {author} {\bibfnamefont {R.~A.}\ \bibnamefont
  {Duine}}\ and\ \bibinfo {author} {\bibfnamefont {H.~T.~C.}\ \bibnamefont
  {Stoof}},\ }\href {\doibase 10.1103/PhysRevLett.86.2204} {\bibfield
  {journal} {\bibinfo  {journal} {Phys. Rev. Lett.}\ }\textbf {\bibinfo
  {volume} {86}},\ \bibinfo {pages} {2204} (\bibinfo {year} {2001})},\ \Eprint
  {http://arxiv.org/abs/cond-mat/0007320} {arXiv:cond-mat/0007320} \BibitemShut
  {NoStop}%
\bibitem [{\citenamefont {Khamehchi}\ \emph {et~al.}(2017)\citenamefont
  {Khamehchi}, \citenamefont {Hossain}, \citenamefont {Mossman}, \citenamefont
  {Zhang}, \citenamefont {Busch}, \citenamefont {Forbes},\ and\ \citenamefont
  {Engels}}]{Khamehchi2017}%
  \BibitemOpen
  \bibfield  {author} {\bibinfo {author} {\bibfnamefont {M.~A.}\ \bibnamefont
  {Khamehchi}}, \bibinfo {author} {\bibfnamefont {K.}~\bibnamefont {Hossain}},
  \bibinfo {author} {\bibfnamefont {M.~E.}\ \bibnamefont {Mossman}}, \bibinfo
  {author} {\bibfnamefont {Y.}~\bibnamefont {Zhang}}, \bibinfo {author}
  {\bibfnamefont {T.}~\bibnamefont {Busch}}, \bibinfo {author} {\bibfnamefont
  {M.~M.}\ \bibnamefont {Forbes}}, \ and\ \bibinfo {author} {\bibfnamefont
  {P.}~\bibnamefont {Engels}},\ }\href {\doibase
  10.1103/PhysRevLett.118.155301} {\bibfield  {journal} {\bibinfo  {journal}
  {Phys. Rev. Lett.}\ }\textbf {\bibinfo {volume} {118}},\ \bibinfo {pages}
  {155301} (\bibinfo {year} {2017})}\BibitemShut {NoStop}%
\bibitem [{\citenamefont {Hawking}(1992)}]{Hawking1991}%
  \BibitemOpen
  \bibfield  {author} {\bibinfo {author} {\bibfnamefont {S.~W.}\ \bibnamefont
  {Hawking}},\ }\href {\doibase 10.1103/PhysRevD.46.603} {\bibfield  {journal}
  {\bibinfo  {journal} {Phys. Rev. D}\ }\textbf {\bibinfo {volume} {46}},\
  \bibinfo {pages} {603} (\bibinfo {year} {1992})}\BibitemShut {NoStop}%
\bibitem [{\citenamefont {Alcubierre}(1994)}]{Alcubierre1994}%
  \BibitemOpen
  \bibfield  {author} {\bibinfo {author} {\bibfnamefont {M.}~\bibnamefont
  {Alcubierre}},\ }\href {\doibase 10.1088/0264-9381/11/5/001} {\bibfield
  {journal} {\bibinfo  {journal} {Class. Quant. Grav.}\ }\textbf {\bibinfo
  {volume} {11}},\ \bibinfo {pages} {L73} (\bibinfo {year} {1994})},\ \Eprint
  {http://arxiv.org/abs/gr-qc/0009013} {arXiv:gr-qc/0009013} \BibitemShut
  {NoStop}%
\bibitem [{\citenamefont {Rolnick}(1969)}]{Rolnick1969}%
  \BibitemOpen
  \bibfield  {author} {\bibinfo {author} {\bibfnamefont {W.~B.}\ \bibnamefont
  {Rolnick}},\ }\href {\doibase 10.1103/PhysRev.183.1105} {\bibfield  {journal}
  {\bibinfo  {journal} {Phys. Rev.}\ }\textbf {\bibinfo {volume} {183}},\
  \bibinfo {pages} {1105} (\bibinfo {year} {1969})}\BibitemShut {NoStop}%
\bibitem [{\citenamefont {O'Neill}(1983)}]{Oneill1983}%
  \BibitemOpen
  \bibfield  {author} {\bibinfo {author} {\bibfnamefont {B.}~\bibnamefont
  {O'Neill}},\ }\href
  {http://www.amazon.com/Semi-Riemannian-Geometry-Applications-Relativity-Mathematics/dp/0125267401}
  {\emph {\bibinfo {title} {Semi-Riemannian Geometry With Applications to
  Relativity, 103, Volume 103 (Pure and Applied Mathematics)}}}\ (\bibinfo
  {publisher} {Academic Press},\ \bibinfo {year} {1983})\BibitemShut {NoStop}%
\bibitem [{\citenamefont {Beem}\ \emph {et~al.}(1996)\citenamefont {Beem},
  \citenamefont {Ehrlich},\ and\ \citenamefont {Easley}}]{Beem1996}%
  \BibitemOpen
  \bibfield  {author} {\bibinfo {author} {\bibfnamefont {J.}~\bibnamefont
  {Beem}}, \bibinfo {author} {\bibfnamefont {P.}~\bibnamefont {Ehrlich}}, \
  and\ \bibinfo {author} {\bibfnamefont {K.}~\bibnamefont {Easley}},\ }\href
  {https://books.google.es/books?id=N9v-F0VRQR0C} {\emph {\bibinfo {title}
  {Global Lorentzian Geometry, Second Edition}}},\ Chapman \& Hall/CRC Pure and
  Applied Mathematics\ (\bibinfo  {publisher} {Taylor \& Francis},\ \bibinfo
  {year} {1996})\BibitemShut {NoStop}%
\bibitem [{\citenamefont {S\'anchez}(2021)}]{SanchezCaja2021}%
  \BibitemOpen
  \bibfield  {author} {\bibinfo {author} {\bibfnamefont {M.}~\bibnamefont
  {S\'anchez}},\ }\href@noop {} {\enquote {\bibinfo {title} {Some
  counterexamples about globally hyperbolic spacetimes},}\ } (\bibinfo {year}
  {2021}),\ \Eprint {http://arxiv.org/abs/2110.13672} {arXiv:2110.13672
  [gr-qc]} \BibitemShut {NoStop}%
\bibitem [{\citenamefont {{Misner}}(1967)}]{Misner1967}%
  \BibitemOpen
  \bibfield  {author} {\bibinfo {author} {\bibfnamefont {C.~W.}\ \bibnamefont
  {{Misner}}},\ }\enquote {\bibinfo {title} {{Taub-Nut Space as a
  Counterexample to almost anything}},}\ in\ \href@noop {} {\emph {\bibinfo
  {booktitle} {Relativity Theory and Astrophysics. Vol.1: Relativity and
  Cosmology}}},\ Vol.~\bibinfo {volume} {8},\ \bibinfo {editor} {edited by\
  \bibinfo {editor} {\bibfnamefont {J.}~\bibnamefont {{Ehlers}}}}\ (\bibinfo
  {publisher} {Rhode Island: American Mathematical Society},\ \bibinfo {year}
  {1967})\ p.\ \bibinfo {pages} {160}\BibitemShut {NoStop}%
\bibitem [{\citenamefont {Novikov}(1992)}]{Novikov1989b}%
  \BibitemOpen
  \bibfield  {author} {\bibinfo {author} {\bibfnamefont {I.~D.}\ \bibnamefont
  {Novikov}},\ }\href {\doibase 10.1103/PhysRevD.45.1989} {\bibfield  {journal}
  {\bibinfo  {journal} {Phys. Rev. D}\ }\textbf {\bibinfo {volume} {45}},\
  \bibinfo {pages} {1989} (\bibinfo {year} {1992})}\BibitemShut {NoStop}%
\bibitem [{\citenamefont {Echeverria}\ \emph {et~al.}(1991)\citenamefont
  {Echeverria}, \citenamefont {Klinkhammer},\ and\ \citenamefont
  {Thorne}}]{Echeverria1991}%
  \BibitemOpen
  \bibfield  {author} {\bibinfo {author} {\bibfnamefont {F.}~\bibnamefont
  {Echeverria}}, \bibinfo {author} {\bibfnamefont {G.}~\bibnamefont
  {Klinkhammer}}, \ and\ \bibinfo {author} {\bibfnamefont {K.~S.}\ \bibnamefont
  {Thorne}},\ }\href {\doibase 10.1103/PhysRevD.44.1077} {\bibfield  {journal}
  {\bibinfo  {journal} {Phys. Rev. D}\ }\textbf {\bibinfo {volume} {44}},\
  \bibinfo {pages} {1077} (\bibinfo {year} {1991})}\BibitemShut {NoStop}%
\bibitem [{\citenamefont {Friedman}\ \emph {et~al.}(1990)\citenamefont
  {Friedman}, \citenamefont {Morris}, \citenamefont {Novikov}, \citenamefont
  {Echeverria}, \citenamefont {Klinkhammer}, \citenamefont {Thorne},\ and\
  \citenamefont {Yurtsever}}]{Friedman1990b}%
  \BibitemOpen
  \bibfield  {author} {\bibinfo {author} {\bibfnamefont {J.}~\bibnamefont
  {Friedman}}, \bibinfo {author} {\bibfnamefont {M.~S.}\ \bibnamefont
  {Morris}}, \bibinfo {author} {\bibfnamefont {I.~D.}\ \bibnamefont {Novikov}},
  \bibinfo {author} {\bibfnamefont {F.}~\bibnamefont {Echeverria}}, \bibinfo
  {author} {\bibfnamefont {G.}~\bibnamefont {Klinkhammer}}, \bibinfo {author}
  {\bibfnamefont {K.~S.}\ \bibnamefont {Thorne}}, \ and\ \bibinfo {author}
  {\bibfnamefont {U.}~\bibnamefont {Yurtsever}},\ }\href {\doibase
  10.1103/PhysRevD.42.1915} {\bibfield  {journal} {\bibinfo  {journal} {Phys.
  Rev. D}\ }\textbf {\bibinfo {volume} {42}},\ \bibinfo {pages} {1915}
  (\bibinfo {year} {1990})}\BibitemShut {NoStop}%
\bibitem [{\citenamefont {Volovik}(2016)}]{Volovik2016}%
  \BibitemOpen
  \bibfield  {author} {\bibinfo {author} {\bibfnamefont {G.~E.}\ \bibnamefont
  {Volovik}},\ }\href {\doibase 10.1134/S0021364016210050} {\bibfield
  {journal} {\bibinfo  {journal} {Pisma Zh. Eksp. Teor. Fiz.}\ }\textbf
  {\bibinfo {volume} {104}},\ \bibinfo {pages} {660} (\bibinfo {year}
  {2016})},\ \Eprint {http://arxiv.org/abs/1610.00521} {arXiv:1610.00521
  [cond-mat.other]} \BibitemShut {NoStop}%
\bibitem [{\citenamefont {Nissinen}\ and\ \citenamefont
  {Volovik}(2017)}]{Volovik2017}%
  \BibitemOpen
  \bibfield  {author} {\bibinfo {author} {\bibfnamefont {J.}~\bibnamefont
  {Nissinen}}\ and\ \bibinfo {author} {\bibfnamefont {G.~E.}\ \bibnamefont
  {Volovik}},\ }\href {\doibase 10.1134/S0021364017070013} {\bibfield
  {journal} {\bibinfo  {journal} {JETP Lett.}\ }\textbf {\bibinfo {volume}
  {105}},\ \bibinfo {pages} {447} (\bibinfo {year} {2017})},\ \Eprint
  {http://arxiv.org/abs/1702.04624} {arXiv:1702.04624 [cond-mat.str-el]}
  \BibitemShut {NoStop}%
\bibitem [{\citenamefont {Sab\'\i{}n}(2016)}]{Sabin2016}%
  \BibitemOpen
  \bibfield  {author} {\bibinfo {author} {\bibfnamefont {C.}~\bibnamefont
  {Sab\'\i{}n}},\ }\href {\doibase 10.1103/PhysRevD.94.081501} {\bibfield
  {journal} {\bibinfo  {journal} {Phys. Rev. D}\ }\textbf {\bibinfo {volume}
  {94}},\ \bibinfo {pages} {081501} (\bibinfo {year} {2016})},\ \Eprint
  {http://arxiv.org/abs/1603.00639} {arXiv:1603.00639 [quant-ph]} \BibitemShut
  {NoStop}%
\bibitem [{\citenamefont {Mart\'\i{}n-V\'azquez}\ and\ \citenamefont
  {Sab\'\i{}n}(2020)}]{Sabin2018}%
  \BibitemOpen
  \bibfield  {author} {\bibinfo {author} {\bibfnamefont {G.}~\bibnamefont
  {Mart\'\i{}n-V\'azquez}}\ and\ \bibinfo {author} {\bibfnamefont
  {C.}~\bibnamefont {Sab\'\i{}n}},\ }\href {\doibase 10.1088/1361-6382/ab5f3f}
  {\bibfield  {journal} {\bibinfo  {journal} {Class. Quant. Grav.}\ }\textbf
  {\bibinfo {volume} {37}},\ \bibinfo {pages} {045013} (\bibinfo {year}
  {2020})},\ \Eprint {http://arxiv.org/abs/1810.05124} {arXiv:1810.05124
  [quant-ph]} \BibitemShut {NoStop}%
\bibitem [{\citenamefont {Penrose}\ \emph {et~al.}(1993)\citenamefont
  {Penrose}, \citenamefont {Sorkin},\ and\ \citenamefont
  {Woolgar}}]{Penrose1993}%
  \BibitemOpen
  \bibfield  {author} {\bibinfo {author} {\bibfnamefont {R.}~\bibnamefont
  {Penrose}}, \bibinfo {author} {\bibfnamefont {R.~D.}\ \bibnamefont {Sorkin}},
  \ and\ \bibinfo {author} {\bibfnamefont {E.}~\bibnamefont {Woolgar}},\
  }\href@noop {} {\enquote {\bibinfo {title} {{A Positive mass theorem based on
  the focusing and retardation of null geodesics}},}\ } (\bibinfo {year}
  {1993}),\ \Eprint {http://arxiv.org/abs/gr-qc/9301015} {arXiv:gr-qc/9301015}
  \BibitemShut {NoStop}%
\bibitem [{\citenamefont {Olum}(1998)}]{Olum1998}%
  \BibitemOpen
  \bibfield  {author} {\bibinfo {author} {\bibfnamefont {K.~D.}\ \bibnamefont
  {Olum}},\ }\href {\doibase 10.1103/PhysRevLett.81.3567} {\bibfield  {journal}
  {\bibinfo  {journal} {Phys. Rev. Lett.}\ }\textbf {\bibinfo {volume} {81}},\
  \bibinfo {pages} {3567} (\bibinfo {year} {1998})},\ \Eprint
  {http://arxiv.org/abs/gr-qc/9805003} {arXiv:gr-qc/9805003} \BibitemShut
  {NoStop}%
\bibitem [{\citenamefont {Visser}\ \emph {et~al.}(2000)\citenamefont {Visser},
  \citenamefont {Bassett},\ and\ \citenamefont {Liberati}}]{Visser1998}%
  \BibitemOpen
  \bibfield  {author} {\bibinfo {author} {\bibfnamefont {M.}~\bibnamefont
  {Visser}}, \bibinfo {author} {\bibfnamefont {B.}~\bibnamefont {Bassett}}, \
  and\ \bibinfo {author} {\bibfnamefont {S.}~\bibnamefont {Liberati}},\ }\href
  {\doibase 10.1016/S0920-5632(00)00782-9} {\bibfield  {journal} {\bibinfo
  {journal} {Nucl. Phys. B Proc. Suppl.}\ }\textbf {\bibinfo {volume} {88}},\
  \bibinfo {pages} {267} (\bibinfo {year} {2000})},\ \Eprint
  {http://arxiv.org/abs/gr-qc/9810026} {arXiv:gr-qc/9810026} \BibitemShut
  {NoStop}%
\bibitem [{\citenamefont {Visser}\ \emph {et~al.}(1999)\citenamefont {Visser},
  \citenamefont {Bassett},\ and\ \citenamefont {Liberati}}]{Visser1999}%
  \BibitemOpen
  \bibfield  {author} {\bibinfo {author} {\bibfnamefont {M.}~\bibnamefont
  {Visser}}, \bibinfo {author} {\bibfnamefont {B.}~\bibnamefont {Bassett}}, \
  and\ \bibinfo {author} {\bibfnamefont {S.}~\bibnamefont {Liberati}},\ }\href
  {\doibase 10.1063/1.1301601} {\bibfield  {journal} {\bibinfo  {journal} {AIP
  Conf. Proc.}\ }\textbf {\bibinfo {volume} {493}},\ \bibinfo {pages} {301}
  (\bibinfo {year} {1999})},\ \Eprint {http://arxiv.org/abs/gr-qc/9908023}
  {arXiv:gr-qc/9908023} \BibitemShut {NoStop}%
\bibitem [{\citenamefont {Gao}\ and\ \citenamefont {Wald}(2000)}]{Gao2000}%
  \BibitemOpen
  \bibfield  {author} {\bibinfo {author} {\bibfnamefont {S.}~\bibnamefont
  {Gao}}\ and\ \bibinfo {author} {\bibfnamefont {R.~M.}\ \bibnamefont {Wald}},\
  }\href {\doibase 10.1088/0264-9381/17/24/305} {\bibfield  {journal} {\bibinfo
   {journal} {Class. Quant. Grav.}\ }\textbf {\bibinfo {volume} {17}},\
  \bibinfo {pages} {4999} (\bibinfo {year} {2000})},\ \Eprint
  {http://arxiv.org/abs/gr-qc/0007021} {arXiv:gr-qc/0007021} \BibitemShut
  {NoStop}%
\bibitem [{\citenamefont {Kim}\ and\ \citenamefont {Thorne}(1991)}]{Kim1991}%
  \BibitemOpen
  \bibfield  {author} {\bibinfo {author} {\bibfnamefont {S.~W.}\ \bibnamefont
  {Kim}}\ and\ \bibinfo {author} {\bibfnamefont {K.~S.}\ \bibnamefont
  {Thorne}},\ }\href {\doibase 10.1103/PhysRevD.43.3929} {\bibfield  {journal}
  {\bibinfo  {journal} {Phys. Rev. D}\ }\textbf {\bibinfo {volume} {43}},\
  \bibinfo {pages} {3929} (\bibinfo {year} {1991})}\BibitemShut {NoStop}%
\bibitem [{\citenamefont {Visser}(2002)}]{Visser2002}%
  \BibitemOpen
  \bibfield  {author} {\bibinfo {author} {\bibfnamefont {M.}~\bibnamefont
  {Visser}},\ }in\ \href@noop {} {\emph {\bibinfo {booktitle} {{Workshop on
  Conference on the Future of Theoretical Physics and Cosmology in Honor of
  Steven Hawking's 60th Birthday}}}}\ (\bibinfo {year} {2002})\ \Eprint
  {http://arxiv.org/abs/gr-qc/0204022} {arXiv:gr-qc/0204022} \BibitemShut
  {NoStop}%
\bibitem [{\citenamefont {Barcel\'o}\ \emph {et~al.}(2021)\citenamefont
  {Barcel\'o}, \citenamefont {Carballo-Rubio}, \citenamefont {Garay},\ and\
  \citenamefont {Garc\'\i{}a-Moreno}}]{Barcelo2021}%
  \BibitemOpen
  \bibfield  {author} {\bibinfo {author} {\bibfnamefont {C.}~\bibnamefont
  {Barcel\'o}}, \bibinfo {author} {\bibfnamefont {R.}~\bibnamefont
  {Carballo-Rubio}}, \bibinfo {author} {\bibfnamefont {L.~J.}\ \bibnamefont
  {Garay}}, \ and\ \bibinfo {author} {\bibfnamefont {G.}~\bibnamefont
  {Garc\'\i{}a-Moreno}},\ }\href {\doibase 10.3390/app11188763} {\bibfield
  {journal} {\bibinfo  {journal} {Appl. Sciences}\ }\textbf {\bibinfo {volume}
  {11}},\ \bibinfo {pages} {8763} (\bibinfo {year} {2021})},\ \Eprint
  {http://arxiv.org/abs/2108.06582} {arXiv:2108.06582 [gr-qc]} \BibitemShut
  {NoStop}%
\bibitem [{\citenamefont {Rosen}(1940{\natexlab{a}})}]{Rosen1940a}%
  \BibitemOpen
  \bibfield  {author} {\bibinfo {author} {\bibfnamefont {N.}~\bibnamefont
  {Rosen}},\ }\href {\doibase 10.1103/PhysRev.57.147} {\bibfield  {journal}
  {\bibinfo  {journal} {Phys. Rev.}\ }\textbf {\bibinfo {volume} {57}},\
  \bibinfo {pages} {147} (\bibinfo {year} {1940}{\natexlab{a}})}\BibitemShut
  {NoStop}%
\bibitem [{\citenamefont {Rosen}(1940{\natexlab{b}})}]{Rosen1940b}%
  \BibitemOpen
  \bibfield  {author} {\bibinfo {author} {\bibfnamefont {N.}~\bibnamefont
  {Rosen}},\ }\href {\doibase 10.1103/PhysRev.57.150} {\bibfield  {journal}
  {\bibinfo  {journal} {Phys. Rev.}\ }\textbf {\bibinfo {volume} {57}},\
  \bibinfo {pages} {150} (\bibinfo {year} {1940}{\natexlab{b}})}\BibitemShut
  {NoStop}%
\bibitem [{\citenamefont {Alvarez}\ \emph {et~al.}(2006)\citenamefont
  {Alvarez}, \citenamefont {Blas}, \citenamefont {Garriga},\ and\ \citenamefont
  {Verdaguer}}]{Alvarez2006}%
  \BibitemOpen
  \bibfield  {author} {\bibinfo {author} {\bibfnamefont {E.}~\bibnamefont
  {Alvarez}}, \bibinfo {author} {\bibfnamefont {D.}~\bibnamefont {Blas}},
  \bibinfo {author} {\bibfnamefont {J.}~\bibnamefont {Garriga}}, \ and\
  \bibinfo {author} {\bibfnamefont {E.}~\bibnamefont {Verdaguer}},\ }\href
  {\doibase 10.1016/j.nuclphysb.2006.08.003} {\bibfield  {journal} {\bibinfo
  {journal} {Nucl. Phys. B}\ }\textbf {\bibinfo {volume} {756}},\ \bibinfo
  {pages} {148} (\bibinfo {year} {2006})},\ \Eprint
  {http://arxiv.org/abs/hep-th/0606019} {arXiv:hep-th/0606019} \BibitemShut
  {NoStop}%
\bibitem [{\citenamefont {Emparan}\ and\ \citenamefont
  {Toma\v{s}evi\'c}(2021)}]{Emparan2021a}%
  \BibitemOpen
  \bibfield  {author} {\bibinfo {author} {\bibfnamefont {R.}~\bibnamefont
  {Emparan}}\ and\ \bibinfo {author} {\bibfnamefont {M.}~\bibnamefont
  {Toma\v{s}evi\'c}},\ }\href@noop {} {\enquote {\bibinfo {title} {{Holography
  of time machines}},}\ } (\bibinfo {year} {2021}),\ \Eprint
  {http://arxiv.org/abs/2107.14200} {arXiv:2107.14200 [hep-th]} \BibitemShut
  {NoStop}%
\bibitem [{\citenamefont {Carlip}(2014)}]{Carlip2013}%
  \BibitemOpen
  \bibfield  {author} {\bibinfo {author} {\bibfnamefont {S.}~\bibnamefont
  {Carlip}},\ }\href {\doibase 10.1016/j.shpsb.2012.11.002} {\bibfield
  {journal} {\bibinfo  {journal} {Stud. Hist. Phil. Sci. B}\ }\textbf {\bibinfo
  {volume} {46}},\ \bibinfo {pages} {200} (\bibinfo {year} {2014})},\ \Eprint
  {http://arxiv.org/abs/1207.2504} {arXiv:1207.2504 [gr-qc]} \BibitemShut
  {NoStop}%
\bibitem [{\citenamefont {Dolgov}\ and\ \citenamefont
  {Novikov}(1998)}]{Dolgov1998}%
  \BibitemOpen
  \bibfield  {author} {\bibinfo {author} {\bibfnamefont {A.~D.}\ \bibnamefont
  {Dolgov}}\ and\ \bibinfo {author} {\bibfnamefont {I.~D.}\ \bibnamefont
  {Novikov}},\ }\href {\doibase 10.1016/S0370-2693(98)01223-4} {\bibfield
  {journal} {\bibinfo  {journal} {Phys. Lett. B}\ }\textbf {\bibinfo {volume}
  {442}},\ \bibinfo {pages} {82} (\bibinfo {year} {1998})},\ \Eprint
  {http://arxiv.org/abs/gr-qc/9807067} {arXiv:gr-qc/9807067} \BibitemShut
  {NoStop}%
\bibitem [{\citenamefont {Adams}\ \emph {et~al.}(2006)\citenamefont {Adams},
  \citenamefont {Arkani-Hamed}, \citenamefont {Dubovsky}, \citenamefont
  {Nicolis},\ and\ \citenamefont {Rattazzi}}]{Adams2006}%
  \BibitemOpen
  \bibfield  {author} {\bibinfo {author} {\bibfnamefont {A.}~\bibnamefont
  {Adams}}, \bibinfo {author} {\bibfnamefont {N.}~\bibnamefont {Arkani-Hamed}},
  \bibinfo {author} {\bibfnamefont {S.}~\bibnamefont {Dubovsky}}, \bibinfo
  {author} {\bibfnamefont {A.}~\bibnamefont {Nicolis}}, \ and\ \bibinfo
  {author} {\bibfnamefont {R.}~\bibnamefont {Rattazzi}},\ }\href {\doibase
  10.1088/1126-6708/2006/10/014} {\bibfield  {journal} {\bibinfo  {journal}
  {JHEP}\ }\textbf {\bibinfo {volume} {10}},\ \bibinfo {pages} {014} (\bibinfo
  {year} {2006})},\ \Eprint {http://arxiv.org/abs/hep-th/0602178}
  {arXiv:hep-th/0602178} \BibitemShut {NoStop}%
\bibitem [{\citenamefont {Camanho}\ \emph {et~al.}(2016)\citenamefont
  {Camanho}, \citenamefont {Edelstein}, \citenamefont {Maldacena},\ and\
  \citenamefont {Zhiboedov}}]{Camanho2014}%
  \BibitemOpen
  \bibfield  {author} {\bibinfo {author} {\bibfnamefont {X.~O.}\ \bibnamefont
  {Camanho}}, \bibinfo {author} {\bibfnamefont {J.~D.}\ \bibnamefont
  {Edelstein}}, \bibinfo {author} {\bibfnamefont {J.}~\bibnamefont
  {Maldacena}}, \ and\ \bibinfo {author} {\bibfnamefont {A.}~\bibnamefont
  {Zhiboedov}},\ }\href {\doibase 10.1007/JHEP02(2016)020} {\bibfield
  {journal} {\bibinfo  {journal} {JHEP}\ }\textbf {\bibinfo {volume} {02}},\
  \bibinfo {pages} {020} (\bibinfo {year} {2016})},\ \Eprint
  {http://arxiv.org/abs/1407.5597} {arXiv:1407.5597 [hep-th]} \BibitemShut
  {NoStop}%
\bibitem [{\citenamefont {Liberati}\ \emph {et~al.}(2002)\citenamefont
  {Liberati}, \citenamefont {Sonego},\ and\ \citenamefont
  {Visser}}]{Liberati2001}%
  \BibitemOpen
  \bibfield  {author} {\bibinfo {author} {\bibfnamefont {S.}~\bibnamefont
  {Liberati}}, \bibinfo {author} {\bibfnamefont {S.}~\bibnamefont {Sonego}}, \
  and\ \bibinfo {author} {\bibfnamefont {M.}~\bibnamefont {Visser}},\ }\href
  {\doibase 10.1006/aphy.2002.6233} {\bibfield  {journal} {\bibinfo  {journal}
  {Annals Phys.}\ }\textbf {\bibinfo {volume} {298}},\ \bibinfo {pages} {167}
  (\bibinfo {year} {2002})},\ \Eprint {http://arxiv.org/abs/gr-qc/0107091}
  {arXiv:gr-qc/0107091} \BibitemShut {NoStop}%
\bibitem [{\citenamefont {Babichev}\ \emph {et~al.}(2008)\citenamefont
  {Babichev}, \citenamefont {Mukhanov},\ and\ \citenamefont
  {Vikman}}]{Babichev2007}%
  \BibitemOpen
  \bibfield  {author} {\bibinfo {author} {\bibfnamefont {E.}~\bibnamefont
  {Babichev}}, \bibinfo {author} {\bibfnamefont {V.}~\bibnamefont {Mukhanov}},
  \ and\ \bibinfo {author} {\bibfnamefont {A.}~\bibnamefont {Vikman}},\ }\href
  {\doibase 10.1088/1126-6708/2008/02/101} {\bibfield  {journal} {\bibinfo
  {journal} {JHEP}\ }\textbf {\bibinfo {volume} {02}},\ \bibinfo {pages} {101}
  (\bibinfo {year} {2008})},\ \Eprint {http://arxiv.org/abs/0708.0561}
  {arXiv:0708.0561 [hep-th]} \BibitemShut {NoStop}%
\bibitem [{\citenamefont {Geroch}(2011)}]{Geroch2010}%
  \BibitemOpen
  \bibfield  {author} {\bibinfo {author} {\bibfnamefont {R.}~\bibnamefont
  {Geroch}},\ }\href@noop {} {\bibfield  {journal} {\bibinfo  {journal} {AMS/IP
  Stud. Adv. Math.}\ }\textbf {\bibinfo {volume} {49}},\ \bibinfo {pages} {59}
  (\bibinfo {year} {2011})},\ \Eprint {http://arxiv.org/abs/1005.1614}
  {arXiv:1005.1614 [gr-qc]} \BibitemShut {NoStop}%
\end{thebibliography}%

\end{document}